\documentclass[review]{elsarticle}

\usepackage{lineno,hyperref}
\modulolinenumbers[5]

\usepackage{xcolor}
\usepackage{graphicx}
\graphicspath{{./img/}}
\DeclareGraphicsExtensions{.pdf}
\usepackage{subfig}

\usepackage{amsmath,amssymb,amsfonts}
\usepackage{algorithmicx,algpseudocode}

\journal{Journal on Parallel and Distributed Computing}

\newcommand{\dflys}{Dragonflies}
\newcommand{\dfly}{Dragonfly}
\newcommand{\ib}{IB}
\newcommand{\ibl}{InfiniBand}

\begin{document}
\begin{frontmatter}

\title{Leveraging InfiniBand Controller to Configure Deadlock-Free Routing Engines for \dflys{}}

\author{German Maglione-Mathey, Jesus Escudero-Sahuquillo, Pedro Javier Garcia, Francisco J. Quiles\fnref{uclmemail}}
\address{Departamento de Sistemas Informáticos, Universidad de Castilla-La Mancha, Spain}
\fntext[uclmemail]{german.maglione@dsi.uclm.es, jesus.escudero@uclm.es, pedrojavier.garcia@uclm.es, francisco.quiles@uclm.es}

\author{Eitan Zahavi\fnref{mnxemail}}
\address{Mellanox Technologies, Israel}
\fntext[mnxemail]{eitan@mellanox.com}

\begin{abstract}
The \dfly{} topology is currently one of the most popular network topologies in high-performance parallel systems.
The interconnection networks of many of these systems are built from components based on the \ibl{} specification.
However, due to some constraints in this specification, the available versions of the \ibl{} network controller (OpenSM) do not include
routing engines based on some popular deadlock-free routing algorithms proposed theoretically for \dflys{},
such as the one proposed by Kim and Dally based on Virtual-Channel shifting.
In this paper we propose a straightforward method to integrate this routing algorithm in OpenSM as a routing engine,
explaining in detail the configuration required to support it.
We also provide experiment results, obtained both from a real \ibl-based cluster and from simulation, to validate
the new routing engine and to compare its performance and requirements against other routing engines currently available in OpenSM.
\end{abstract}


\end{frontmatter}


\section{Motivation}
\label{s_motivation}

High-Performance Computing (HPC) systems, as well as Datacenters, consist of a high number of computing and storage endnodes interconnected by means of a network.
Given the increasingly growing communication requirements of the services and applications supported by these systems,
the performance demanded to the interconnection network (mainly high throughput and low latency) augments accordingly.
Hence, it is essential to optimize the design and configuration of the interconnection network in order to achieve the desired performance required by the system services and applications.

One of the most important design aspects of any interconnection network is its topology, i.e. the pattern followed to interconnect the endnodes of the system.
In that sense, \dfly{} topologies \cite{kim_technology-driven_2008}
have become very popular in the last years because of their low diameter, path diversity, high bisection bandwidth, and good performance/cost ratio.
In addition, they offer a good scalability due to their hierarchical structure.
Thanks to these advantages, \dfly{} topologies have been configured in several real systems such as the IBM PERCS \cite{percs_2010}, the Cray XC series \cite{faanes_cray_2012}
or the Niagara (the Canada's fastest supercomputer) \cite{niagara}, which implements a \dfly$+$ architecture \cite{dflyplus2017}.

However, the interconnection pattern of \dfly{} topologies contains physical cycles that may lead to cyclic channel dependencies and so to traffic deadlocks.
When a packet  is allowed by the routing algorithm to traverse  two channels $c_n$ and $c_m$, in that order, to reach its destination, this creates a \emph{direct dependency} from $c_n$ to $c_m$. The set of all \emph{direct dependencies} that packets create, starting from all sources to all destinations, forms a \emph{channel dependency graph} \cite{dally87_vc}, where each vertex represents a channel and an edge represents a dependency. \emph{Cyclic dependencies} are formed between two or more channels that depend on each other, either directly or indirectly.
This situation may lead to a cycle of packets waiting for one another to release resources (i.e. channels) in a deadlock configuration \cite{dally_principles_2003}, and hence remain blocked indefinitely.
Hence it is mandatory that the routing algorithms used in \dflys{} are designed to guarantee deadlock freedom.
For that purpose, the routing algorithms, either deterministic or adaptive, usually considered as suitable for \dflys{} prevent
potential cyclic dependencies among channels \cite{dally87_vc} by following one out of three potential approaches.
The first approach uses Virtual Channels (VCs) \cite{dally87_vc} as escape ways, i.e. the VC assigned to a packet
may vary along their route so that no cyclic channel dependencies appear in the network for a given VC.
In the second approach, the routes that follows the packets in the network are grouped into disjoint sets, and each set is mapped to an exclusive VC, all this according to some algorithm that guarantees that
not all the routes that form any possible cycle belong to the same set (i.e. that not all of the routes forming a cycle share the same VC).
This second approach is in general known as ``layered routing'' \cite{skeie2002_lash}.
The third approach restricts routes so that the allowed ones never form a cycle.

The use in \dflys{} of routing algorithms based on the first approach was proposed at the same time as this topology by Kim and Dally \cite{kim_technology-driven_2008}.
Specifically, a deterministic minimal routing algorithm that requires 2 VCs to provide deadlock freedom was proposed, as well as an oblivious
non-minimal routing algorithm that requires 3 VCs, the latter being an adaptation of the Valiant's routing algorithm \cite{valiant_1981}.
In addition, it was also suggested the use of both minimal and non-minimal routes according to the adaptive routing algorithm proposed in \cite{kim_technology-driven_2008, singh-thesis2005},
which requires 3 VCs to provide deadlock freedom.
However, these algorithms, similarly to others providing deadlock freedom through the same approach (e.g. the ones proposed in \cite{mgarcia-icpp13, faizian-hoti16, won2015}),
are not easy to implement in real systems due to the requirement of ``shifting'' the VC assigned to some packets.
In particular, this ``VC shifting'' is difficult to achieve in networks built from components based on the InfiniBand architecture \cite{IBA2015} (see Section \ref{s_background_ibrouting}),
that is currently one of the most widely used network technology in HPC systems \cite{top500_list}.
In that sense, although VCs can be emulated through the InfiniBand Virtual Lanes (VLs), some restrictions in the InfiniBand specification  \cite{IBA2015}
(see Section \ref{s_problem_statement}) make it complex to shift the VL assigned to a packet along its route, as required by the routing algorithms that follow the first approach for deadlocks prevention.

For that reason, the deadlock-free routing algorithms that have been implemented as routing engines in the \ibl{} control software (OpenSM \cite{opensm}) follow mainly the second or third approaches mentioned above.
The available routing engines suitable for \dflys{} that follow the second approach (i.e. layered routing) are LASH \citep{skeie2002_lash}, DFSSSP \cite{domke2011_dfsssp} and D3R \cite{maglione2018_d3r},
the former two being actually topology agnostic algorithms while the latter having been specially designed for \dflys{}.
Regarding the third approach, the classical, topology-agnostic Up/Down algorithm is also available as a routing engine (UPDN) in OpenSM \cite{sancho2001_updn}, which can be also used in \dflys{}.
As far as we know, the only deadlock-free routing algorithm following the first approach (i.e. requiring VL shifting) and suitable for \dflys{} that has been implemented as a \ib{} routing engine is DF-DN \cite{schneider2016_dfdn}.
This routing engine actually was proposed in general for low-diameter topologies, and it is based on a non-uniform configuration of the tables that map the packets
Service Level (SL, see Section \ref{s_background_ibrouting}) to VLs, which indeed introduces some degree of complexity in the network configuration.

It is worth clarifying that all the routing engines mentioned in the former paragraph are deterministic, since implementing adaptive routing algorithms in
\ibl{} presents very difficult or even unsurmountable problems.
It is worth clarifying also that
DFSSSP and LASH routing engines are strictly minimal-path and UPDN may provide non-minimal routes.
Nevertheless, D3R is minimal if we consider a \dfly{} group (see Section \ref{s_background_dfly}) as a virtual switch,
therefore D3R minimizes the number of traversed global channels (see Section \ref{s_background_dfly_routing}).
This is also true for the minimal routing algorithm for \dflys{} \cite{kim_technology-driven_2008} (see Section \ref{s_background_dfly_routing}).

In summary, when it comes to configure a \dfly{} topology in a real \ibl-based cluster, the administrator has to choose basically among a not many available deadlock-free,
deterministic routing engines.
However, there are not many studies comparing the performance of these routing engines which could help network administrator in choosing the most suitable one for the specific \dfly{} network under configuration.
In that sense, in \cite{schneider2016_dfdn} the requirements of some of these routing engines are analyzed in terms of routing time and number of required VLs, but no performance measurements are provided.
In \cite{maglione2018_d3r}, both a comparison of the requirements (in terms of number of required VLs) and a performance comparison (based on both simulation experiments and results
from a real \ibl-based cluster) are provided, but not all the routing engines available for \dflys{} are considered.

In order to make easier an optimal configuration of \ibl-based \dfly{} networks, in this paper we first enlarge the set of deadlock-free routing engines available for \dflys{} by proposing a method to implement in
OpenSM the deterministic, minimal routing algorithm proposed originally in \cite{kim_technology-driven_2008}, which, as far as we know, had not been implemented yet as a routing engine.
Second, we analyze the requirements, pros and cons of all the available routing engines in this set, including the new one, considering the impact of several
switch features and constraints in their implementation and performance.
Third, we provide a thorough performance comparison of these routing engines, based on both simulation experiments and results from a real \ibl-based cluster.
For this performance comparison we consider several traffic scenarios, including real applications, the results showing that in many scenarios
the newly implemented routing engine outperforms the other ones while requiring fewer VLs.

The rest of the paper is organized as follows.
Section \ref{s_background} describes the background details of \dfly{} topologies \ibl-based networks and routing engines for \ibl-based clusters.
Section \ref{s_problem_statement} points out the implementation problems of the routing algorithms that are the base of our proposal.
Section \ref{s_implementation} covers the implementation details of the mentioned routing as a routing engine in the \ibl{} network controller.
In Section \ref{s_evaluation} the deadlock-free routing engines available for \dflys{} are analyzed and evaluated based on experiment results.
Finally, in Section \ref{s_conclusions} some conclusions are drawn.

\section{Background}
\label{s_background}

\subsection{\dfly{} Topology}
\label{s_background_dfly}

\begin{figure}[!htb]
	\centering
	\includegraphics[width=0.8\columnwidth]{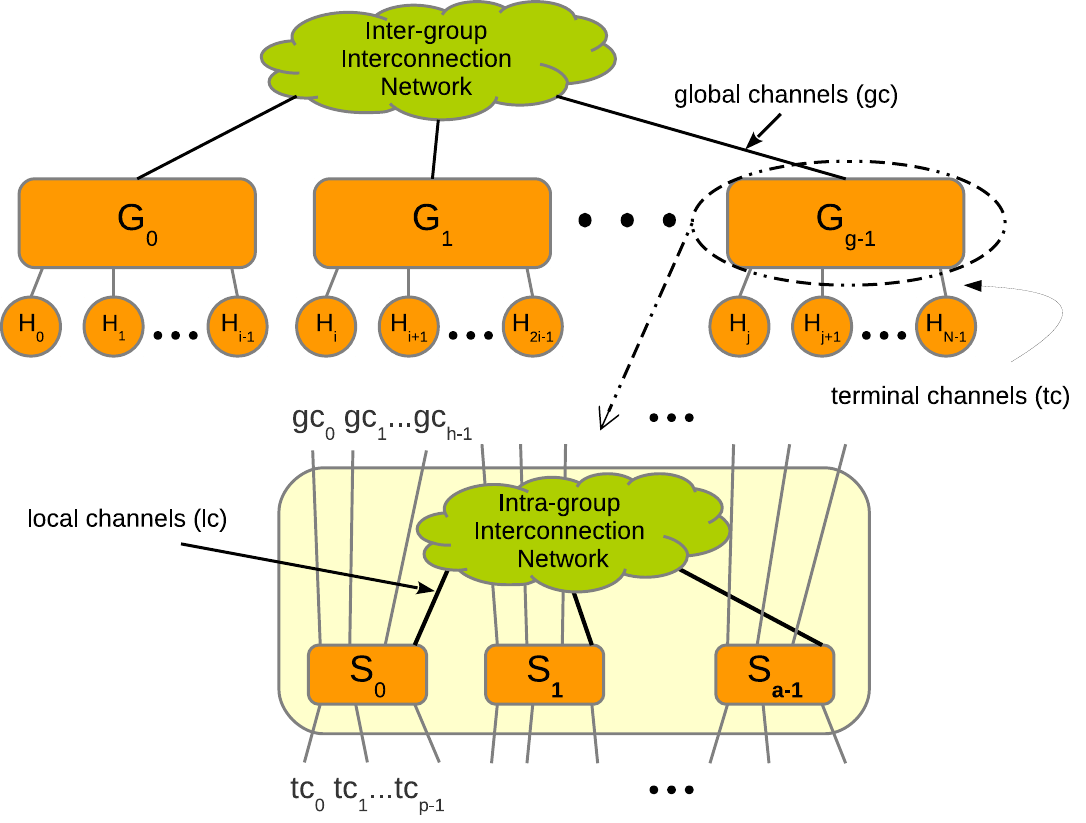}
	\caption{Generic \dfly{} connection pattern.}
	\label{f:dflyschem}
\end{figure}

The \dfly{} \cite{kim_technology-driven_2008} is one of the most popular topologies nowadays, due to its interesting properties \cite{camarero2014_hamm_dfly}.
Figure \ref{f:dflyschem} shows a generic \dfly{} topology, where switches are organized in two hierarchical levels.
In the first level, switches and endnodes into the same group are connected via local channels, forming the intra-group network.
In the second level, the groups are connected by means of global channels that form the inter-group network.
A \dfly{} network can be defined by three parameters:

\begin{itemize}
 \item $a$ switches in each group.
 \item $p$ endnodes connected to each switch.
 \item $h$ links at each switch used to connect to other groups (i.e. the global channels).
\end{itemize}

Based on these parameters a group contains $a$ switches which are interconnected via local channels, and altogether can be considered as a virtual high-radix switch.
Each group is connected to the other groups by means of $a\times h$ global channels.
Both the intra- and inter-group topologies are not tied to any particular connection pattern.
For instance, a fully-connected \dfly{} network assumes a direct link between any pair of switches within the same group (i.e. the intra-group level),
and a direct link between any pair of groups (i.e. the inter-group level).
Fully-connected \dflys{} can interconnect $N = ap(ah + 1)$ endnodes by using $g = ah + 1$ groups. 
In order to balance channel load on load-balanced traffic, fully-connected \dflys{} should be built so that the values of parameters $a$, $h$, and $p$ fulfill the equation $a = 2h = 2p$.
This pattern is the one recommended in the original paper where the \dfly{} topology was proposed \cite{kim_technology-driven_2008}.

\dfly{} networks are cost-effective, since the number of elements required to interconnect large networks is lower than those required by other topologies.
Besides, among the interesting properties of \dfly{} networks, it is worth mentioning the low-diameter (i.e. the average number of hops required to route packets
from a source endnode to a destination endnode) and path diversity (i.e. the different routes that can be followed by packets when communicating two endnodes).

\subsubsection{\dfly{} Routing}
\label{s_background_dfly_routing}

\begin{figure}[!tb]
	\centering
	\includegraphics[width=0.7\columnwidth]{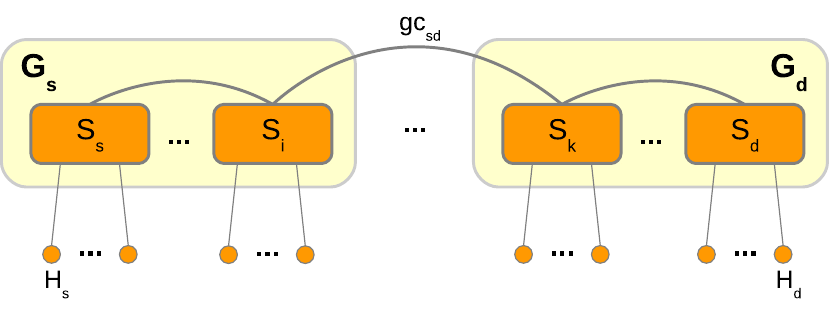}
	\caption{Inter-group path in a \dfly{} topology.}
	\label{f:inter-group-path}
\end{figure}

A deadlock-free minimal routing algorithm for \dflys{} was proposed in the original paper where this topology was defined \cite{kim_technology-driven_2008}.
Figure \ref{f:inter-group-path} shows a minimal routing in a \dfly{}, from source node $H_s$ attached to switch $S_s$ in group $G_s$ to destination node $H_d$ attached to switch $S_d$ in group $G_d$
that traverses a single global channel $gc_{sd}$. The pseudo-code defining this routing algorithm can be seen in Figure \ref{f:dfly_ruting_alg}.
This algorithm, hereinafter referred to as DLA, requires $2$ Virtual Channels (VCs) \cite{dally87_vc} to prevent deadlocks.
Basically, cyclic dependencies are avoided by shifting the VC of the packets when they are going to traverse a local (i.e. intra-group) channel after traversing a global (i.e. inter-group) one.

As mentioned earlier, in order to balance channel load under uniform traffic, a fully-connected \dfly{} should be built using $a = 2h = 2p$.
Analyzing the relationship that $a$, $h$ and $p$ must satisfy to balance channels load requires to compute the number of flows (i.e. source-destination pairs) traversing each channel, assuming unidirectional channels.
A fully-connected \dfly{} has three types of channels: terminal channels ($tc$), local channels ($lc$) and global channels ($gc$) (see Figure \ref{f:dflyschem}).
Assuming a minimal routing, the number of flows traversing a $tc$ is: 
\begin{equation}
 f_t = 1 \times (N - 1) = ap(ah + 1) - 1
\end{equation}

A $gc$ is used only to communicate a pair of groups,
thus the number of flows traversing a $gc$ is:
\begin{equation}
f_g = (ap)^2
\label{e:global_flows}
\end{equation}

Three different types of flows traverse a $lc$ of a group:
\begin{enumerate}
\item Flows starting and ending in that group: $p^2$.
\item Flows starting in that group and ending in a different group: $ahp^2$.
\item Flows starting in a different group and ending in that group: $ahp^2$.
\end{enumerate}

Therefore, the total number of flows traversing a $lc$ is equal to:
\begin{equation}
f_l = p^2 + ahp^2 + ahp^2
\label{e:local_flows}
\end{equation}

For $p \ge 2$, the numbers of flows of the different types of channels satisfy $f_g < f_l < f_t$.

\begin{figure}[!htb]
	\centering
	\includegraphics[width=0.8\columnwidth]{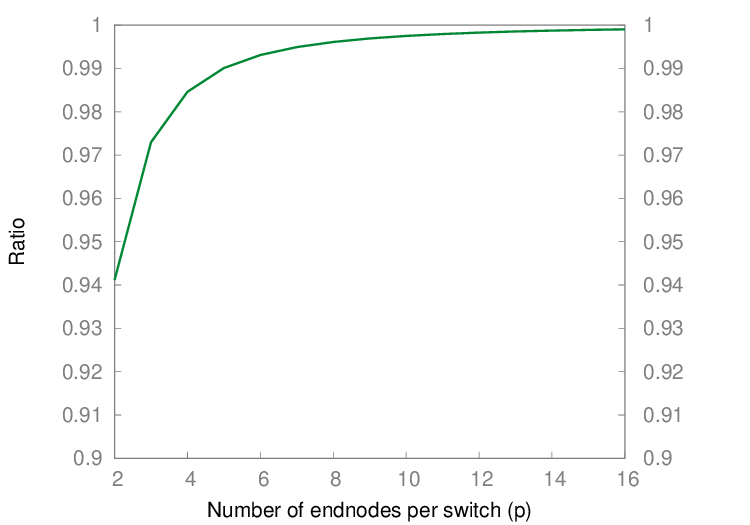}
	\caption{The ratio $\frac{f_g}{f_l}$ is close to $1$ when the network size increases.}
	\label{f:flow_ratio}
\end{figure}

Regarding the load balance of VCs and channels, first let's consider the balance between the two VCs used by the routing algorithm.
In the case of the different types of flows traversing a $lc$ (enumerated above), the first two types are mapped to the same VC, but the last (third) type of flow to a different VC.
Hence, the imbalance between the numbers of flows mapped to these two VCs ($p^2 + ahp^2$ and $ahp^2$, respectively) is driven by the term $p^2$. However,
this difference decreases in relation to the number of flows mapped to each VC when the network size increases, since $a$ increases twice as fast as $p$.
Similarly, the imbalance between the numbers of flows traversing the types of channels $gc$ and $lc$ is driven by the term $p^2$. This can be seen by rearranging Equation \ref{e:local_flows}
as follows  $ f_l = 2ahp^2 + p^2$, and replacing $2h = a$ (as we assume fully-connected \dflys{}), then resulting $ f_l = (ap)^2 + p^2$.
Figure \ref{f:flow_ratio} shows how the ratio $\frac{f_g}{f_l}$ approaches $1$ when $p$ increases, so also the network size.

This minimal routing works well for load-balanced traffic, but results in very poor performance under adversarial traffic patterns \cite{kim_technology-driven_2008}.

\begin{figure}[!htb]
\begin{algorithmic}[0]
       \If{$G_s \ne G_d \And S_s$ is not directly connected to $G_d$}
                \State route to $S_i$ \Comment $S_i$ has a connection ($gc_{sd}$) to $G_d$
       \ElsIf{$G_s \ne G_d$}
                \State route to $S_k$ \Comment $S_k$ is a switch in group $G_d$

       \ElsIf{$S_k \ne S_d$}
                \State route to $S_d$
       \EndIf
\end{algorithmic}
\caption{Pseudo-code of the minimal routing for \dflys{} (DLA).}
\label{f:dfly_ruting_alg}
\end{figure}

As an alternative to minimal routing, the original \dfly{} paper also proposes an adaptation of the non-minimal Valiant's routing algorithm \cite{valiant_1981} to \dflys{},
so that packets are routed first to a randomly-chosen intermediate group, then to their actual destination group.
In this way, a uniform, random distribution of traffic emerges even if the original traffic pattern is adversarial-like, balancing traffic load on local and global links
so that the performance degradation that this pattern may generate is prevented \cite{kim_technology-driven_2008}.
However, this is achieved at the expense of longer paths, which may increase packet latency, and at the cost of requiring $3$ VCs to prevent deadlocks (packet VC must be shifted after every inter-group hop).
Moreover, the use of longer routes required by Valiant's routing algorithm leads to approximately a doubling of the traffic load that is traversing the netwok with respect to the minimal-path routing \cite{dally_principles_2003, prisacari-hpdc2014},
and this may subsequently hasten to an unnecessary performance degradation for uniform-like traffic patterns.

As both minimal and non-minimal routing may produce a performance degradation under different traffic patterns, the use of adaptive routing is also suggested in the original \dfly{} paper,
specifically the \emph{Universal Globally-Adaptive Load-Balanced} (UGAL) algorithm \cite{singh-thesis2005}.
UGAL balances traffic among minimal and non-minimal paths on a packet-by-packet basis, the select of the path being made by using 
the queue occupancy level and hop count information to select the path with minimum delay.
For the \dfly{} topology, two versions of UGAL were initially proposed: UGAL-L, that uses local queue occupancy information at the current node,
and UGAL-G, that uses that information from all the global channels (although the latter is considered ideal and unfeasible in practice \cite{kim_technology-driven_2008}).
Both versions of UGAL require $3$ VCs to provide deadlock freedom in \dflys{}.
By contrast, there are proposals of deadlock-free adaptive routing algorithms for \dflys{} that do not need 
VCs to prevent deadlocks \cite{mgarcia2012_ofar, dxiang2016}, although in practice they require other additional network resources.
In general, deadlock-free adaptive routings for \dflys{} introduce some degree of network complexity and demand a higher number of network resources with respect to deterministic minimal routing.
Moreover, adaptive routing may require packet reordering at the destination endnodes upon out-of-order delivery, then introducing additional latency.
As mentioned above, all these problems of adaptive routing would be introduced unnecessarily in scenarios where minimal routing suffices to achieve good performance.
Moreover, even in the case that adaptive routing is required, deadlock-free minimal routing may be still necessary since it is necessary
to select among minimal and non-minimal paths, as it happens when UGAL is used.
Note that all the routing algorithms proposed in the original \dfly{} paper \cite{kim_technology-driven_2008} require shifting VCs, which makes complex their implementation
in \ib-based networks (further details can be found in Section \ref{s_problem_statement}).
For that reason, other algorithms have been traditionally used to provide deadlock-free routing to \ib-based networks, as explained in Section \ref{s_background_ibrouting}.

\subsection{\ibl{} Architecture Basics}
\label{s_background_iba}

The \ibl{} Architecture (\ib{}) specification \cite{IBA2015} describes a virtual cut-through (VCT) interconnect technology for communicating processing and/or storage nodes (i.e. endnodes).
Emerging applications requiring intensive computing and massive storage have imposed on high-performance computer (HPC) clusters, datacenters and hyperscalers an increased burden,
demanding the interconnection network greater capacity to move data between endnodes, and pushing more functionality down to the network elements.
\ib-based networks offer higher bandwidth and lower latency compared to other network technologies.
Last generation of \ib-based products provides HDR speed (i.e. 200 Gb/s) and 0.6$\mu$s end-to-end latency.
It is expected that these features will improve in the future \cite{iba_roadmap} in order to accomplish with the demands of future applications.

The \ib-based networks are composed of several elements: network interfaces or host channel adapters (HCAs), switches, routers and cables to interconnect the former elements.
The subnet manager (SM) that is also a part of an \ib{} network is a control software entity that discovers and configures the network.
HCAs are in charge of injecting in the network packets generated by the applications.
These packets travel throughout the network, reaching switches and routers via the ports.
Switches and routers have several ports, so that \ib-based networks can be configured with several network topologies, such as Meshes, Tori, Fat-trees or \dflys{}, among others.
Cables connect HCA ports to the switch/router ports.
Moreover, network elements in \ib{} networks have to be identified (addressed) so that routing algorithms can define routes in the topology.
The addressing in \ib{} networks is defined by means of local identifiers (LIDs) that are assigned to HCA and switch/router ports in the network-configuration stage by the SM.
Then, these ports can be identified unequivocally within the network when the topology is discovered, as well as the network elements (i.e. HCAs, switches and routers) that they belong to.

As mentioned above, in \ib-based networks, the configuration stage is orchestrated by the SM, an agent that may be implemented in software or hardware, which discovers all the network elements that form the topology.
More precisely, the SM broadcasts management datagrams (MADs) that are received by the network elements (several MADs can be received by the same network element through several paths in the network).
Then these elements notify that they exist to the SM, and the SM sends them their corresponding LID. 
Obviously, the network topology must offer full connectivity among these network elements so that they can be discovered.
At the same time the SM identifies the network elements, it discovers the network topology by keeping track of the paths followed by MADs when discovering the network elements.
After discovering the topology, the SM applies the selected routing algorithm by populating the Linear Forwarding Tables (LFTs) at switches.
At the end of this stage, the SM is able to provide the network description, containing the LIDs of network elements, the network topology and the routing algorithm information at the switch routing tables.

\begin{figure}[!htb]
	\centering
	\includegraphics[width=0.5\textwidth]{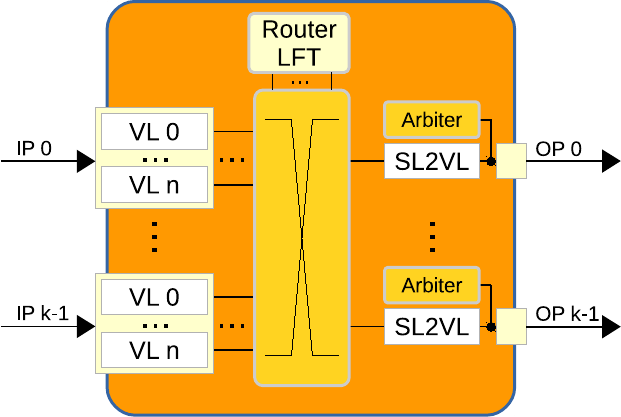}
	\caption{Diagram of the architecture of an \ib-based $k$-port switch.}
	\label{f:ibaswschem}
\end{figure}

In the \ib{} specification, the buffering at HCA and switch ports is channeled through Virtual Lanes (VLs).
Basically the physical links, interconnecting buffers between two ports at different network elements are split into a set of logical VLs sharing the total link bandwidth.
The \ib{} specification defines a credit-based flow-control scheme at VL-level, so that,
VLs are allowed to forward a packet only if there are available credits for the VL assigned to that packet at the next switch (or HCA) port.
Hence, VLs also represent a fraction of the total buffer space at each port.
In \ib-based networks, VLs are assigned to packets based on their Service Level (SL), a numerical identifier that is set to packets at HCAs, prior the injection of the packets in the network.
The SL value cannot be changed once their injection in the network.
Meanwhile, the maximum number of VLs per HCA or switch port defined in the \ib{} specification is $15$, although several manufacturers such as Mellanox limit this number to $9$ VLs
($1$ of them reserved for management purposes).
The number of SLs available in \ib-based networks is limited to $16$.
Figure \ref{f:ibaswschem} shows a diagram of an \ib-based $k$-port switch.

As can be seen in Figure \ref{f:ibaswschem} every switch (and HCA) has a SL-to-VL mapping table (SL2VL) per output port, which is used to assign packets
requesting that output port with a specific VL, based on those packets SLs and on their arrival port (in the case of switches).
For that purpose, each entry of the SL2VL table associates an input port and a SL with a VL.
Hence, a packet can be assigned to different VLs along its route, depending on its SL and on the information of the SL2VL table at each HCA or switch output port crossed by this packet.
In addition to the LFTs, the SM also populates the SL2VL tables both at HCAs and switch ports.

OpenSM \cite{opensm} is the open-source implementation of the SM included in the OpenFabrics Software (OFS) \cite{web_ofa}, a commonly used open-source control-software distribution for \ib-based networks.
OFS is supported by the OpenFabrics Alliance \cite{web_ofa}, which gathers several companies promoting the \ib{} technology.
Basically, OpenSM implements the functionality of the SM, including the implementation of different routing algorithms, referred to as \emph{routing engines}.
In particular, the routing engines are in charge of computing the routing tables and send this information to the switches.

\subsection{Deadlock-free Routing in \ibl-based \dflys{}}
\label{s_background_ibrouting}

As mentioned above, the interconnection pattern of the \dfly{} topologies contains physical cycles that may lead to deadlocks \cite{dally_principles_2003},
thus the routing algorithms used in \dflys{} must be designed to avoid the possibility of such deadlocks to emerge.
Indeed, a deadlock situation (``credit loops'' in the \ibl{} (\ib{}) jargon) can emerge in an \ib-based network,
because cyclic dependencies among channels may appears due to the backpressure over the buffers associated with these channels.

In general, there are three different approaches to design a deadlock-free routing algorithm.
The first approach uses Virtual Channels (VCs) \cite{dally87_vc} as escape ways, where the VC assigned to a packet
may vary along its route.
This approach performs a ``shifting'' of the VC assigned to a packet, so that no cyclic dependencies appear in the network for a given VC.
As mentioned above, shifting VCs is difficult to implement in \ib-based networks (see Section \ref{s_problem_statement} for more details),
hence this approach has been traditionally avoided in real \ib-based systems.
Indeed, as far as we know, besides the implementation of DLA that we propose in this paper, DF-DN \cite{schneider2016_dfdn} is currently the only \ib{} routing engine that uses VC-shifting
(VL-shifting in \ib{} terms), which is restricted to low-diameter \ib{} networks.
It is worth clarifying that DF-DN leaves the calculation of the Linear Forwarding Tables (LFTs) to routing engines that do not
prevent deadlocks, such as SSSP \cite{sssp_hoefler_2009} and MINHOP \cite{opensm}, while it configures the SL2VL tables so that deadlocks are avoided.

In the second approach, known as Layered Routing, the routing engine analyzes the paths that may form a cycle in the channel dependency graph \citep{dally_principles_2003},
and maps some of those paths to different VLs in order to break cyclic channel dependencies.
Note that in this approach VLs are assigned for the entire paths, so that packets cannot change VL in the middle of their route (i.e. there is no VL-shifting).
Hence, this approach has been traditionally the preferred one to design routing engines.
Current routing engines available in \ib{} that are suitable for \dfly{} networks and use Layered Routing, are DFSSSP \cite{domke2011_dfsssp},
LASH \cite{skeie2002_lash} and D3R \cite{maglione2018_d3r}.

The last approach to design deadlock-free routing algorithms is based on restricting routes so that the allowed ones never form a cycle.
OpenSM provides two similar algorithms that follow this approach and can be used in a \dfly{} network:
Up*/Down* \cite{up_down_1991} (UPDN \cite{sancho2001_updn}) and  Down*/Up* \cite{up_down_1991} (DNUP \cite{sancho2001_updn}).

In summary, the available routing engines that are able to make a deadlock-free routing configuration for a \dfly{} network are:
\begin{enumerate}
    \item {\bf DLA}: Our own implementation of the minimal topology-aware routing for \dfly{}, proposed in this paper \cite{kim_technology-driven_2008}.
     It will be explained in detail in Section \ref{s_implementation}.

    \item {\bf D3R}: It is a topology-aware, minimal, deterministic, deadlock-free and scalable routing algorithm, suitable for
    \dfly{} topologies that use a fully-connected pattern in the inter-group network, and any connection pattern in the intra-group network.
    D3R maps each route to a single, specific VL depending on the destination group, and according to a specific order, so that deadlocks are prevented \cite{maglione2018_d3r}.

   \item {\bf UPDN}:  It is a topology-agnostic, deterministic and deadlock free routing engine \cite{sancho2001_updn} that
   builds the spanning tree of the interconnection network, then removing (i.e. restricting) paths in order to break cyclic dependencies.  
   However, UPDN provides non-minimal routes when applied to \dfly{} topologies.
   DNUP is similar to UPDN.
		
   \item {\bf LASH}: Is an unicast topology-agnostic routing engine for \ib{}  \cite{skeie2002_lash}.
   It provide deadlock-free minimal-path routing while distributing paths among VLs (Layers).
   Depending on the network size, the number of required VLs varies.

   \item {\bf DF-DN}: It imposes a total order on virtual channels where packets are mapped along their routes, and increments the VL at every hop (which is known as VL-hopping), thus the diameter of the network
   is an upper bound for the required number of VLs \cite{schneider2016_dfdn}.
   Note that this requires to configure the SL2VL tables so that the VL of packets moving between two specific switches is increased.
   However, this approach presents some limitations (see Section \ref{s_problem_statement}).
   DF-DN leaves the calculation of the LFTs to SSSP and MINHOP, two minimal-path topology-agnostic deadlock-free routing engines that optimize link utilization, balancing the number of routes per link.
   
   \item {\bf DFSSSP}: It is a topology-agnostic deadlock-free minimal-path routing engine \cite{domke2011_dfsssp} available in OpenSM.
   DFSSSP globally balances the number of routes per link in order to optimize link utilization.
   As this routing engine is not optimized for \dflys{}, it ends up using too many VLs to break cycles in large networks.
\end{enumerate}

\section{Problem Statement}
\label{s_problem_statement} 

\begin{figure}[!htb]
	\centering
	\includegraphics[width=\columnwidth]{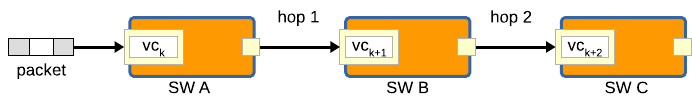}
	\caption{VC-hopping approach to avoid cyclic dependencies.}
	\label{f:vchopping}
\end{figure}

As explained in previous sections, different types of routing algorithms proposed for \dfly{} networks, either minimal or non-minimal
avoid cyclic channel dependencies by shifting the VC of the packets along their path.
A generalized form of this approach is shown in Figure \ref{f:vchopping}, where the Virtual Channel (VC) of the packet changes in every hop (VC hopping) \cite{scherson1994},
so that cyclic channel dependencies are avoided.
Note, however, that some routing algorithms such as the minimal one proposed originally for \dflys{} (DLA) requires shifting VLs only once.
Indeed, changing the packets VC at every hop may be a waste of resources that otherwise could be used for other purposes.

\begin{figure}[!htb]
	\centering
	\includegraphics[width=0.7\columnwidth]{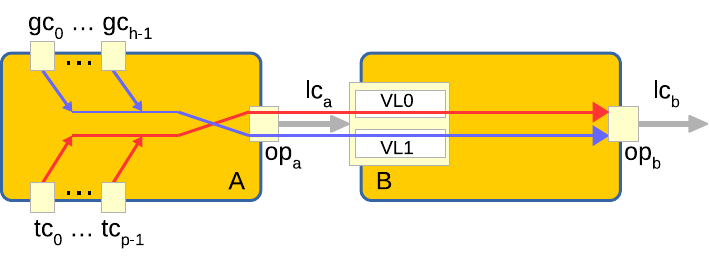}
	\caption{External flows (blue) and local flows (red) inside a \dfly{} group.}
	\label{f:vl-hopping-problem}
\end{figure}

Nevertheless, regardless of the times that packets must shift their VC, the problems to implement VC shifting in \ib-based networks are basically the same.
Indeed, despite the fact that \ib{} Virtual Lanes (VLs) can be thought of as VCs, this approach cannot be used directly in \ib-based networks.
First of all, note that the routing algorithms based on VC shifting require to do it at specific hops along their route, i.e. the VC that 
must be assigned to the packet depends on which hop the packet is performing.
However, there is no ``hop'' information in \ib{} packets.
Moreover, note that the VL assignment is managed through the SL2VL table at each output port, that only use the input port and
Service Level (SL) of a packet to assign the next VL (see Section \ref{s_background_iba}).
Regarding the SL, it cannot contribute in general to identify the hop that the packet is performing, because it cannot be changed once
the packet is injected, and also because the number of available SLs is very reduced.
On the other hand, the input port may help in specific cases to identify the number of hops performed by a packet.
For instance, a packet coming from a port connected to a terminal channel is clearly performing its first hop.
Similarly, a packet coming from a port connected to a global channel has just performed a hop between two \dfly{} groups.
However, in the case of packets coming from ports connected to local channels, we cannot distinguish between packets
originated in a different group and packets whose source is in the current group.
An example of this problem is shown in Figure \ref{f:vl-hopping-problem}.

Specifically, Figure \ref{f:vl-hopping-problem} shows a traffic situation in a \dfly{} group with a non-fully-connected intra-group pattern.
In this case ``external'' packet flows (the blue ones) need to traverse at least two switches (switches A and B) until they
reach their destination switch (not shown in the figure). ``Local'' packet flows (the red ones) follow the same path.

As can be seen, at output port \emph{$op_a$} the information inside the SL2VL table is enough to split into different VLs the flows coming from global channels ($gc_0$ to $gc_{h-1}$)
and the flows coming from terminal channels ($tc_0$ to $tc_{p-1}$).
However, at output port \emph{$op_b$} it is not possible to discern between local (red) and external (blue) flows, because all of them share the same input port (connected to a local channel).
This prevents the correct VL assignment, as external and local flows should be assigned with different VLs to prevent prevent deadlocks.

Taking this problem into account, our \ib-based DLA implementation is limited to \dfly{} networks with a fully-connected pattern in the inter-group and the intra-group network,
because in this way, an external packet flow requires to traverse just one  local channel (lc) inside its destination group before reaching its destination switch.
Therefore, an external flow and a local flow only meet at the first output port that both cross in the group, where is possible to discern between both
flows as they come from local and terminal channels, respectively.
In summary, due to this restriction the input port information is enough to change the VL of the external packet flow as required by the routing algorithm.

\section{DLA Implementation Overview}
\label{s_implementation}

This Section describes the implementation details of the deadlock-free minimal routing for fully-connected \dfly{} networks (DLA),
described in Section \ref{s_background_dfly}, and summarized in Figure \ref{f:dfly_ruting_alg},
as a new routing engine in OpenSM.

According to the routing algorithm, the DLA routing engine uses VLs as escape ways.
i.e. cyclic dependencies are avoided by shifting the VL of the packets when they are going to traverse a local (i.e. intra-group)
channel after traversing a global (i.e. inter-group) one.

Like any other routing engine, DLA operates during the \ib-based network configuration stage.
Specifically, once OpenSM assigns LIDs to all the \ib-based devices (i.e. endnodes and switches),
the DLA routing engine discovers the \dfly{} network topology composed of switches belonging to different groups.
Afterwards, it populates the routing tables (LFTs) and the SL2VL tables.
These basic functionalities are performed in three stages:

\begin{enumerate}
\item \dfly{} group discovery. For this stage we use the calculation of the \emph{closed neighborhood} for each switch \cite{maglione2018_d3r}.
The closed neighborhood of a vertex $v$ in a graph $G$ is the subgraph containing vertex $v$ itself and all vertices adjacent to $v$.

\item Once switches and endnodes are assigned to a \dfly{} group, each switch is visited, and for each target LID,
      a decision is made as to what port should be used to access that LID, according to the algorithm defined in Figure \ref{f:dfly_ruting_alg}.
      This information is used to populate the routing tables (LFTs).
      
\item The last stage involves the SL2VL table calculation for each switch, as explained in the following paragraphs.
\end{enumerate}

For the sake of simplicity we can see all the SL2VL tables at an \ibl{} (\ib{}) switch as one single table.
This table would be indexed by the \emph{output port (op)}, \emph{input port (ip)} and \emph{Service Level (SL)},
to return the corresponding Virtual Lane (VL).
This table can be thought of as a function that assigns a set of input values ($op$, $ip$, $sl$) to a output value ($vl$).

The function shown in Figure \ref{f:sl2vl_function}
is the one applied by the routing engine to configure the SL2VL tables.
Specifically, for any SL, the function returns VL~$1$ when a packet traverses a local (i.e. intra-group) channel (lc)
after traversing a global (i.e. inter-group) channel (gc).
Otherwise, the function returns VL~$0$.

\begin{figure}[!htb]
	\centering
    \[ sl2vl(op, ip, sl) =
      \begin{cases}
        VL~1       & \quad \text{if } op \in \{lc_0 \dots lc_{a-2}\} \land ip \in \{gc_0 \dots gc_{h-1}\}\\
        VL~0       & \quad \text{otherwise}
      \end{cases}
    \]
	\caption{Function used by the DLA routing engine to map packets to VLs.}
	\label{f:sl2vl_function}
\end{figure}

By using this function, the DLA routing engine performs as required by the routing algorithm, hence it is able to avoid deadlocks using only $2$ VLs. Note, however, that this mapping of packets to VLs is independent of the packet's SL, so that optionally a single SL (any) could be used for all the packets for the sake of simplicity.

\begin{figure}[!htb]
	\centering
	\includegraphics[width=0.7\columnwidth]{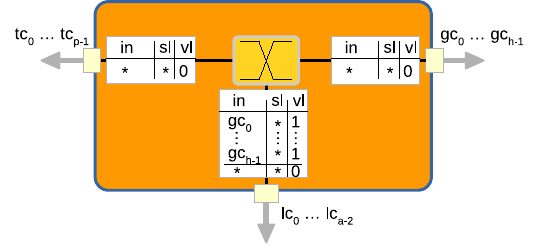}
	\caption{Schematic of an \ib-based switch with SL2VL tables configured according to the DLA routing engine.}
	\label{f:sw_sl2vl}
\end{figure}

Figure \ref{f:sw_sl2vl} shows an schematic of an \ib{} switch with populated SL2VL tables, using the \emph{sl2vl} function shown in Figure \ref{f:sl2vl_function}.
Notice that only packets coming from global to local channels shift from VL~$0$ to VL~$1$. Indeed, an ``asymmetric'' configuration of SL2VL tables results from the function, i.e. different ports get different tables

The members of respectively local and global channels in the expression shown in Figure \ref{f:sl2vl_function} correspond to a
\dfly{} network with a fully-connected pattern within the inter-group and intra-group networks,
as our implementation of DLA is restricted to that pattern as explained in Section \ref{s_problem_statement}.

\section{Routing Engines Evaluation}
\label{s_evaluation}

In this section we analyze and evaluate all the deadlock-free routing engines available for \dfly{} networks (see Section \ref{s_background_ibrouting}),
including our implementation of DLA as a routing engine.
Specifically, we study first the pros and cons of these routing engines taking into account the impact on their performance of several
switch features and constraints in their implementation.
Next, we evaluate the scalability of routing engines based on their requirements and on simulation results.
Finally, we analyze the results obtained from real-traffic workloads run in an \ib-based real cluster.

In more detail, in order to get results for this evaluation we have performed both simulations and experiments with real \ib{} hardware,
using a framework which integrates \ib{} control software, \ib-based hardware and OMNeT++-based simulators \cite{OMNeTManual}.
Basically, in the context of this study we have extended  a previously proposed methodology \cite{maglione2018_d3r, raap-tools},
adding support for managing and using SL2VL tables to the flit-level \ib{} simulator contributed
by Mellanox Technologies\texttrademark \cite{maglione2018_d3r}, and
to an in-house packet-level technology-agnostic interconnection networks simulator \cite{yebenes13_sauron}.

\subsection{Impact of Switch Features on Routing Engines}
\label{s_sim_sw_features}
In this section we analyze the impact of having different VL buffer sizes and the use of Virtual Output Queuing (VOQ)
on the performance of the analyzed routing engines.
In that sense, the \ibl{} (\ib{}) specification requires to provide separate buffering resources at switch and HCA ports,
enough space in each Virtual Lane (VL) for at least one packet, and a separate flow-control plane for data VLs.
However, the VLs maximum buffer size and the implementation of physical VOQ \cite{voq_tamir88}
(i.e. input ports implementing separate queues for each output port) are left to the manufacturer's criteria  \cite{IBA2015},
hence the performance of \ib-based fabric (so networks) may vary, depending on the VOQ implementation.

Indeed, the input buffer size has a considerable influence on the network robustness as more packets can be stored if large buffers are provided, especially when traffic burst or congestion occurs.
Also, it is well known that packets in an input-queued switch (such as \ib-based switches) suffer from poor performance due to Head Of Line (HOL) blocking \cite{hol_karol87}.
The VOQ scheme makes it possible to relay packets at the input port VL that are stored behind blocked packets
due to insufficient space in VL at the next hop, reducing the effects of HOL blocking.

\begin{figure}[!htbp]
        \centering
        \includegraphics[width=\textwidth]{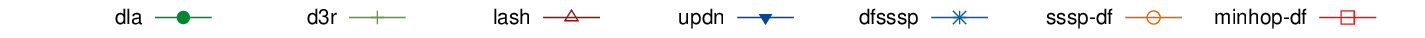}

        \subfloat[1 packet $\times$ VL]{\includegraphics[width=0.45\textwidth]{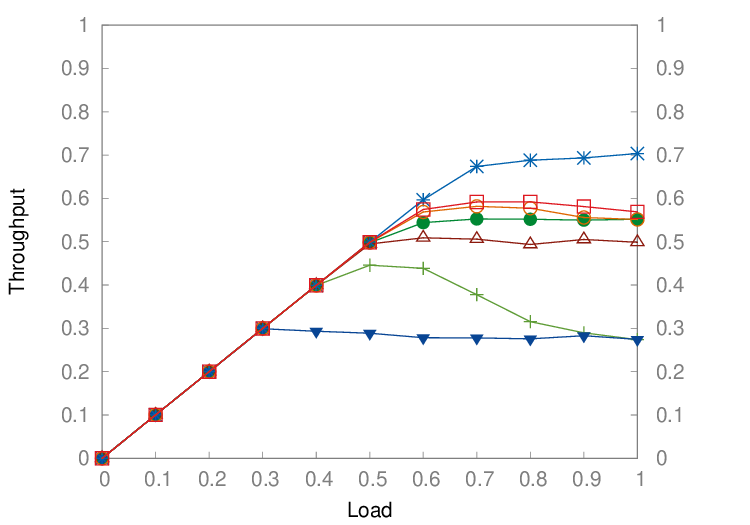}
                 \label{f:r72n-8p-novoq}}
        \subfloat[2 packets $\times$ VL]{\includegraphics[width=0.45\textwidth]{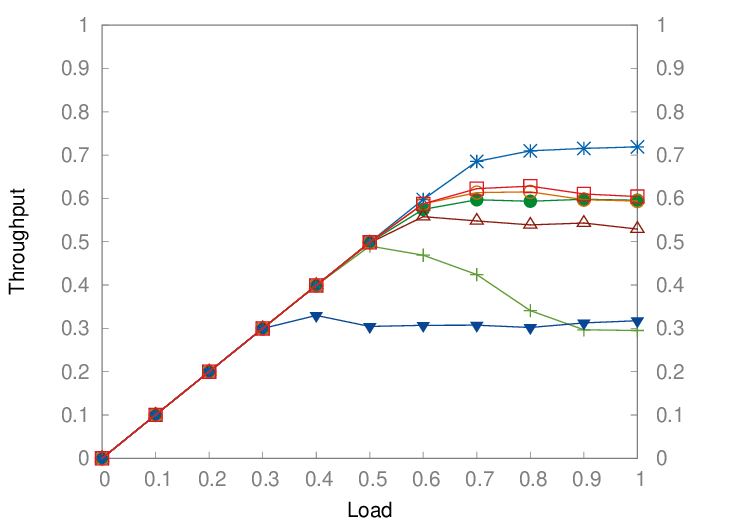}
                 \label{f:r72n-16p-novoq}}
                 
         \subfloat[4 packets $\times$ VL]{\includegraphics[width=0.45\textwidth]{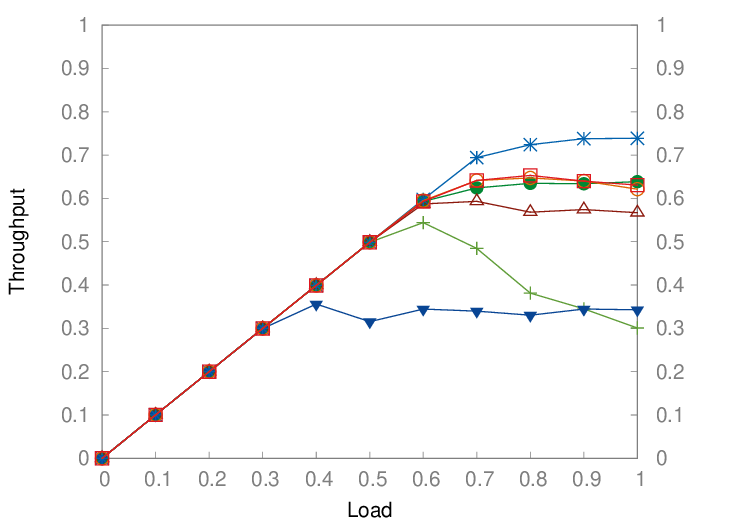}
                 \label{f:r72n-32p-novoq}}
        \subfloat[8 packets $\times$ VL]{\includegraphics[width=0.45\textwidth]{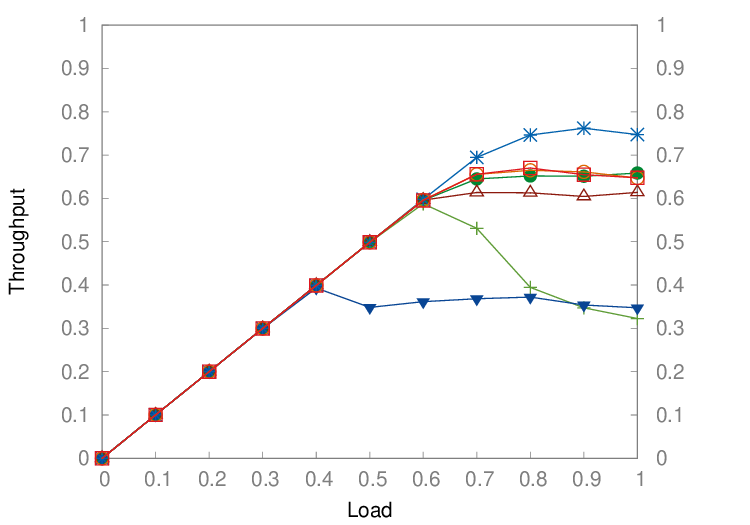}
                 \label{f:r72n-64p-novoq}}
                 
        \subfloat[16 packets $\times$ VL]{\includegraphics[width=0.45\textwidth]{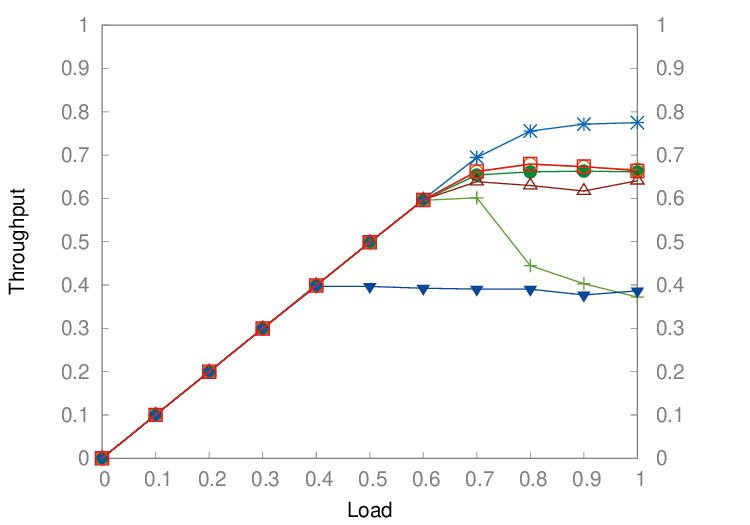}
                 \label{f:r72n-128p-novoq}}
        \subfloat[32 packets $\times$ VL]{\includegraphics[width=0.45\textwidth]{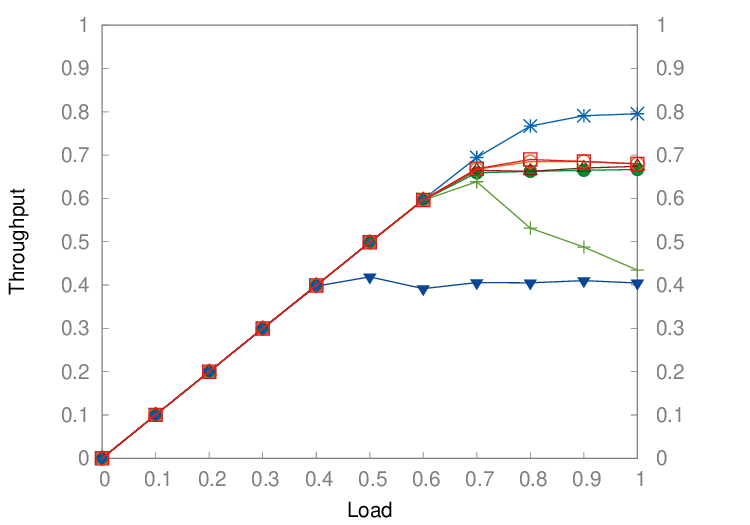}
                 \label{f:r72n-256p-novoq}}

        \caption{72-node \dfly{} without VOQ and different input buffer sizes for dla (1SL, 2VLs), d3r and lash (2SLs, 2VLs), updn (1SL, 1VL), dfsssp (8SLs, 8VLs), sssp-df (3SLs, 3VLs), and minhop-df (4SLs, 3VLs).}

        \label{f:sim-novoq}
\end{figure}

\begin{figure}[!htbp]
        \centering
        \includegraphics[width=\textwidth]{sim-keys}

        \subfloat[1 packet $\times$ VL]{\includegraphics[width=0.45\textwidth]{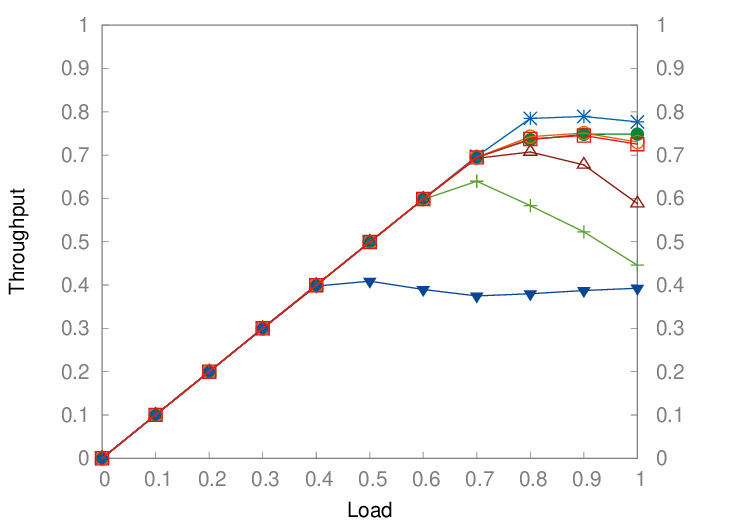}
                 \label{f:r72n-8p-voq}}
        \subfloat[2 packets $\times$ VL]{\includegraphics[width=0.45\textwidth]{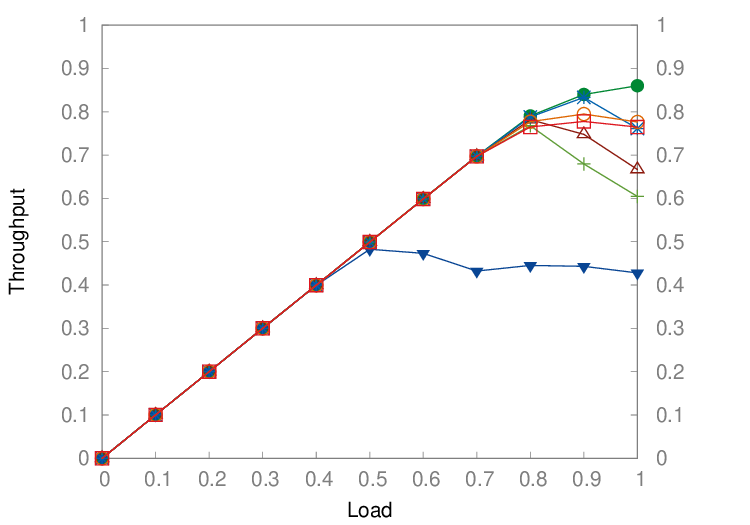}
                 \label{f:r72n-16p-voq}}
                 
        \subfloat[4 packets $\times$ VL]{\includegraphics[width=0.45\textwidth]{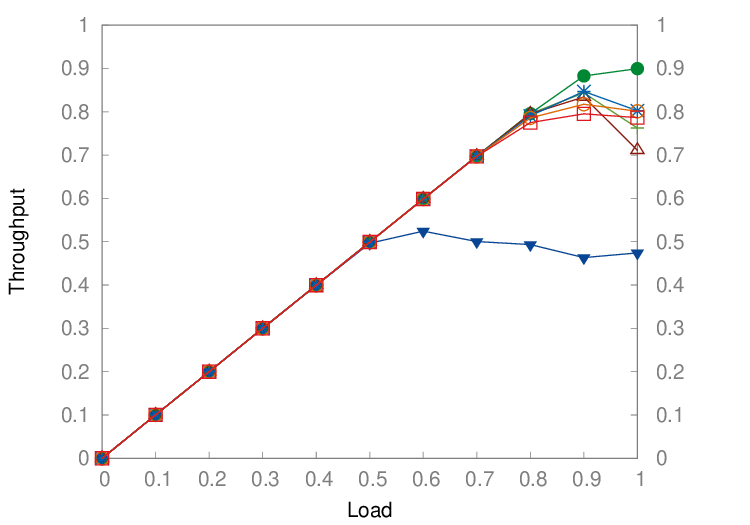}
                 \label{f:r72n-32p-voq}}        
        \subfloat[8 packets $\times$ VL]{\includegraphics[width=0.45\textwidth]{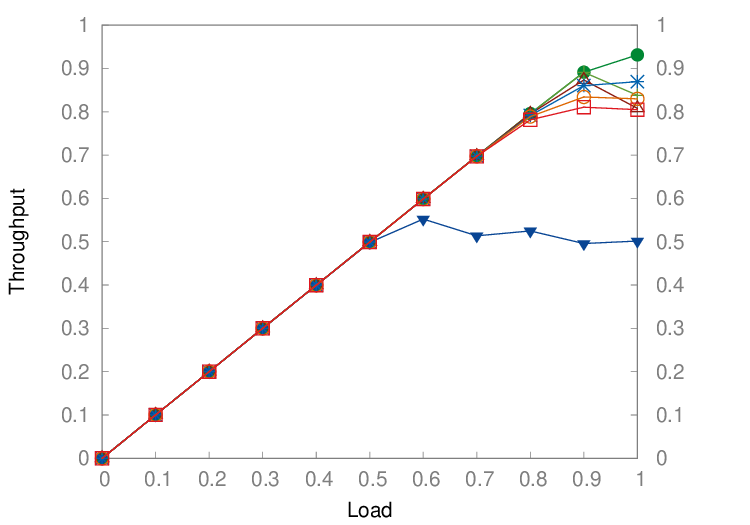}
                 \label{f:r72n-64p-voq}}
                 
        \subfloat[16 packets $\times$ VL]{\includegraphics[width=0.45\textwidth]{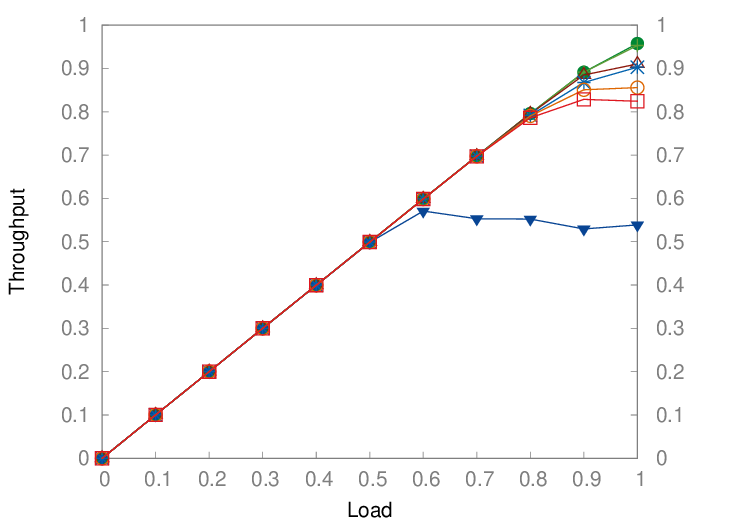}
                 \label{f:r72n-128p-voq}}
        \subfloat[32 packets $\times$ VL]{\includegraphics[width=0.45\textwidth]{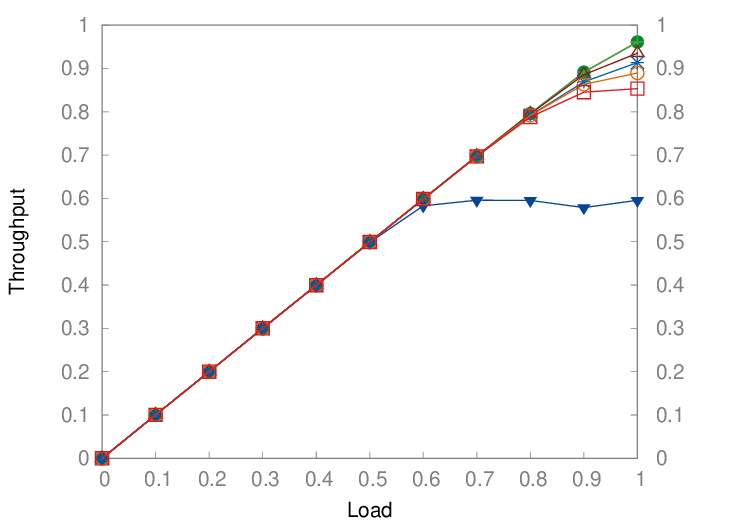}
                 \label{f:r72n-256p-voq}}

        \caption{72-node \dfly{} with VOQ and different input buffer sizes for dla (1SL, 2VLs), d3r and lash (2SLs, 2VLs), updn (1SL, 1VL), dfsssp (8SLs, 8VLs), sssp-df (3SLs, 3VLs), and minhop-df (4SLs, 3VLs).}

        \label{f:sim-voq}
\end{figure}

In order to evaluate the impact of both buffer size and the implementation (or not) of VOQ, we have run simulation experiments
in a \dfly{} topology of fixed size (72 nodes, $a=4$, $h=2$, $p=2$) varying buffer size and modeling switches with or without VOQ.
For that purpose, we have used an in-house developed technology-agnostic packet-level simulator that allows us to test different interconnection
networks configurations and switch features  \cite{yebenes13_sauron}.
The simulator is configured to use a credit-based flow control a VL-level, and static buffering management where each VL has a separate fixed buffer space.
This method prevents VLs from growing beyond a specified size, thereby avoiding buffer hogging \cite{yoshigoe_2007}.
We assume $8$ data VLs and QDR 4x links (i.e. $40$ Gbps of theoretical bandwidth, reduced to $32$ Gbps due to the $8b/10b$ encoding protocol).
We use a MTU size of $4$ KB.
The synthetic traffic pattern used for simulations follows a random uniform distribution of destinations.
Figures \ref{f:sim-novoq} and \ref{f:sim-voq} show the results obtained, specifically the normalized accepted traffic as a function of traffic load.
In that sense, the traffic generation rate have been increased from $0$\% to $100$\% of the maximum link bandwidth ($32$Gbps) in $10\%$ steps.
For each point of load, the results has been computed as the average accepted traffic measured during $5$ms of simulated time,
then normalized against the maximum link bandwidth.

The results for DLA, D3R, LASH, UPDN, DFSSSP, and DF-DN (i.e. SSSP-DF and MINHOP-DF)
for different buffer sizes ($1$, $2$, $4$, $8$, $16$, and $32$ packets per VL) and for switch architectures without VOQ are shown in Figure \ref{f:sim-novoq}.
Figure \ref{f:sim-voq} shows the results for switch architectures with VOQ.
It is worth mentioning that DFSSSP uses $8$ VLs; SSSP-DF and MINHOP-DF $3$ VLs;
DLA, D3R and LASH $2$ VLs; and UPDN uses just one VL (see Table \ref{t:scenarios}).
Therefore, as each VL has a separate fixed buffer space, DFSSSP requires much more overall buffer space than the rest of routing engines.

\begin{figure}[!htbp]
        \centering
        \includegraphics[width=\textwidth]{sim-keys}

        \subfloat[Increasing buffer size without using VOQ]{\includegraphics[width=0.57\textwidth]{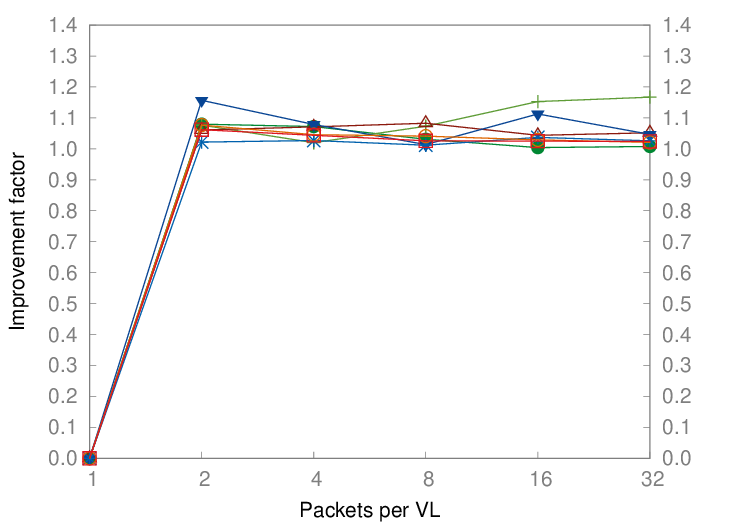}
                 \label{f:r72n-factors-novoq}}
                 
        \subfloat[Increasing buffer size while using VOQ]{\includegraphics[width=0.57\textwidth]{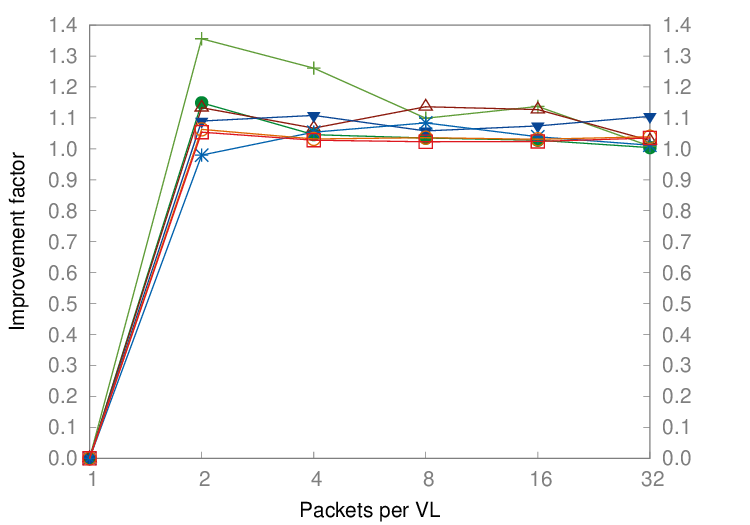}
                 \label{f:r72n-factors-voq}}
                 
        \subfloat[Improvement of VOQ over non-VOQ for the same buffer size]{\includegraphics[width=0.57\textwidth]{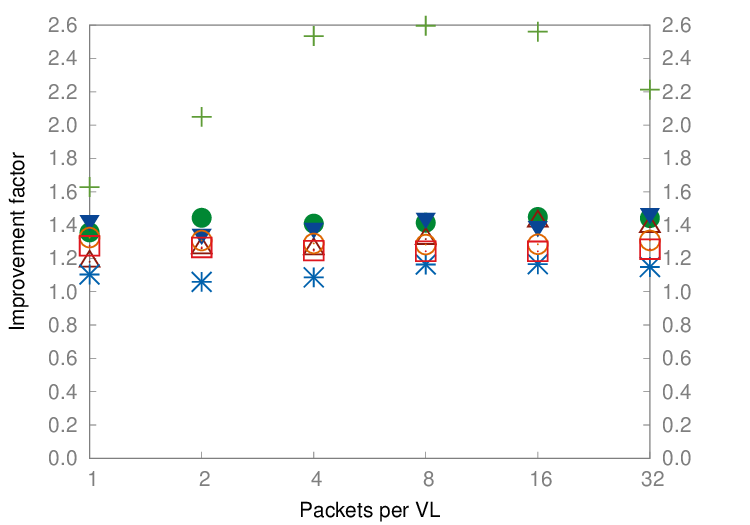}
                 \label{f:r72n-factors}}

        \caption{Impact of different buffer sizes and the use of VOQ for dla (1SL, 2VLs), d3r and lash (2SLs, 2VLs), updn (1SL, 1VL), dfsssp (8SLs, 8VLs), sssp-df (3SLs, 3VLs), and minhop-df (4SLs, 3VLs).}

        \label{f:sim-factors}
\end{figure}

As can be seen, the results show how an increase in the size of the buffer leads to an increase in performance (at a different ratio),
whether VOQ is used (see Figure \ref{f:sim-voq}) or not (see Figure \ref{f:sim-novoq}).
For instance, D3R and UPDN are more responsive to changes in buffer size, especially regarding how much traffic can be absorbed until network reaches the saturation point.
Regarding how much it is worth increasing the size of input buffers, Figures \ref{f:r72n-factors-novoq} and \ref{f:r72n-factors-voq} shows
the improvement factor of each routing engine as the input buffer size is increased from $1$ packet per VL
up to $32$ packets per VL, ether without VOQ (see Figure \ref{f:r72n-factors-novoq})
or with VOQ (see Figure \ref{f:r72n-factors-voq}).
Note that, the overall improvement factor does not grow significantly beyond $16$ packets per VL.

On the other hand, all the routing engines show a considerable performance improvement when VOQ is used.
It is worth noting that DFSSSP shows a stable performance even in the absence of VOQ,
because it uses much more VLs, thus reducing though VLs a greater fraction of HOL blocking \cite{voq-sw_mgomez2003} than the other routing engines.
Indeed, when VOQ is used, DFSSSP performance increases only by a factor of $1.16$, in the best case (see Figure \ref{f:r72n-factors}).
By contrast, the use of VOQ boosts the performance of both DLA and D3R (that use $2$ VLs), so that they outperform DFSSSP.
This is a direct consequence of how these algorithms take advantage of the implementation of VOQ.

\begin{figure}[!htb]
	\centering
	\includegraphics[width=\columnwidth]{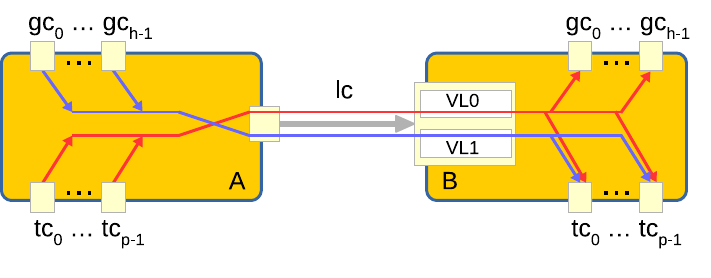}
	\caption{Flows crossing a local channel (intra-group network).}
	\label{f:dla-flows}
\end{figure}

To make this issue clearer, a small subset of a \dfly{} intra-group network with different types of packet flows
is shown in Figure \ref{f:dla-flows}.
Specifically, ``external'' flows (i.e. flows coming from other groups) are shown in blue, and ``local'' flows are shown in red.
In this example, the external flows have their destination in switch B, while local flows have their destination either at switch B or at a group 
connected to switch B.
Therefore, all the flows have to cross the local channel (lc) connecting switches A and B.

In the situation show in the Figure \ref{f:dla-flows}, and according to DLA, local and external flows are assigned with 
different VLs, so that they will not compete for the same buffer space and contention will be reduced.
Furthermore, the use of VOQ allows forwarding packets belonging to a local flow (red flow) that request accessing to global ports $gc_0 \dots gc_{p-1}$
at switch B, even if they share the VL with local flows that are blocked because they request accessing terminal ports that have been granted to external flows.
Overall, this reduces contention on both global and local channels.

By contrast, topology-agnostic routing engines do not distinguish \dfly{} groups (therefore they cannot distinguish between external and local flows),
so that they mix in the same VL locally-generated flows with externally-generated flows, each flow facing different degrees of contention.
In addition, an external flow coming from a global channel $gc_i : 0 \le i < h$ at switch A can be forwarded to any port of that switch, including global channels (except the same port $gc_i$).
Moreover, if the ingress flow is forwarded through a local channel (lc), it may later be forwarded to either terminal or global channels at switch B.
In the case of non-minimal routing engines (such as UPDN), they could keep forwarding the external flow through local channels until destination is reached.
All this increases the number of flows requesting the same output port from different input ports.
Therefore, if an output port becomes saturated, even temporarily, the number of affected flows and input ports is higher for topology agnostic routing engines
than if DLA is used in VOQ architectures.

The improvement factor when using VOQ for the same buffer size is shown in Figure \ref{f:r72n-factors}.
That Figure shows a considerably increase in performance of all the routing engines when using VOQ, going from $1.63$ up to $2.60$ for D3R, to
$1.1$ up to $1.16$ form DFSSSP. The minimum, maximum and median improvement factors for each routing engine are shown in Table \ref{t:voq-ifactors}.

\begin{table}[!hptb]
\caption{\small Minimum, Maximum and Median improvement factors when using VOQ for each routing engine.}
\centering
        \begin{tabular}{|l|c|c|c|} \hline
            Routing engine & Min. & Med. & Max.\\ \hline
            DLA            & 1.356	& 1.428	& 1.448 \\
            D3R            & 1.628	& 2.373	& 2.596 \\
            LASH           & 1.179	& 1.288	& 1.420 \\
            UPDN           & 1.347	& 1.412	& 1.471 \\
            DFSSSP         & 1.059	& 1.126	& 1.165 \\
            SSSP-DF        & 1.283	& 1.298	& 1.324 \\
            MINHOP-DF      & 1.242	& 1.251	& 1.275 \\ \hline
        \end{tabular}
        \label{t:voq-ifactors}
\end{table}

In summary, the DLA routing engine  is suitable for \ib-based \dflys{} from the performance point of view,
provided that there is enough buffer space and a VOQ-like switch architecture is used\footnote{Note that commercial \ib-based components do implement VOQ, such as Mellanox\texttrademark products}.
On the other hand, DFSSSP shows a stable performance (in contrast to other routing engines) regardless of buffer space and whether VOQ is used or not.
However, as DFSSSP uses all the VLs available in comercial components, it cannot be further improved by implementing queuing schemes \cite{dbbm_duato2004, h2lq_yebenes2015}
or Quality of Service (QoS) support.

\subsection{Scalability Analysis}
\label{s_sim-experiments}

In this Section DLA, D3R, LASH, UPDN, DFSSSP, and DF-DN (i.e. SSSP-DF and MINHOP-DF) are evaluated for configurations of \ib-based \dflys{} with different sizes based on simulation results.
For that purpose, we have extended the \ib-specific flit-level simulator contributed by Mellanox\texttrademark \cite{maglione2018_d3r},
implementing the \emph{Service Level to Virtual Lane} (SL2VL) tables.
This simulator uses static buffering management for VLs and implements Virtual Output Queuing (VOQ).
The simulator is configured as follows: $8$ data VLs, QDR 4x links, MTU size of $4$ KB, and buffer size of $16$ packets per VL.
Traffic injection is given by the following synthetic patterns:

\begin{enumerate}
\item \emph{random uniform}: destinations are randomly chosen uniformly among all the endnodes.
                             We sample the network throughput during $5$ms simulated time (after $1$ms of warm-up).
                             
\item \emph{6-point 3D stencil}: it reflects nearest-neighbor communication pattern in real-world applications.
                       We sample the network throughput during $5$ms simulated time (after $1$ms of warm-up).
                       
\item \emph{hot-spot}: some percentage of the endnodes generate traffic directed to a small fraction of endnodes, while the rest of endnodes generate random uniform traffic.
                       We randomly select $h_s = \lfloor 0.06 \times N \rfloor$ (i.e. 6\% of the total endnodes $N$)
                       of endnodes to generate hot-spot traffic at $100$\% of link bandwidth, addressed to $h_d = \lfloor \log_2(h_s) \rfloor$ randomly selected destinations.
                       we sample the network throughput during $1$ms while the congested traffic is being transmitted, after  $1$ms of warm-up,
                       only for those endnodes not receiving hot-spot traffic.
\end{enumerate}

For the first two traffic patterns, the traffic generation rate has been increased from $0$\% to $100$\% of the maximum effective link bandwidth for QDR 4x (i.e. $32$Gbps) in $10$\% steps.
This is also the case for the endnodes in the third (hot-spot) communication pattern that do not generate hot-spot traffic.

Table \ref{t:scenarios} shows the different balanced \dfly{} configurations (see Section \ref{s_background_dfly}) used in the experiments,
as well as the resources (i.e. SLs and VLs) required by each routing engine.
All these network configurations assume a fully-connected intra- and inter-group network.

\begin{table}[!hptb]
\caption{\small Number of SLs and VLs required in fully-connected balanced \dfly{} networks of different sizes.}
\centering
\resizebox{\columnwidth}{!}{%
    \begin{tabular}{|l|c|c|c|c|c|c|c|c|} \hline
    {\# of Nodes } & \multicolumn{2}{c|}{72} & \multicolumn{2}{c|}{342} & \multicolumn{2}{c|}{1056} & \multicolumn{2}{c|}{2550} \\ \hline
    Routing engine & \# SLs & \# VLs         & \# SLs & \# VLs          & \# SLs & \# VLs           & \# SLs & \# VLs \\ \hline
    DLA            &   1    &       2        &   1    &         2       &   1    &       2          &   1    & 2 \\
    D3R            &   2    &       2        &   2    &         2       &   2    &       2          &   2    & 2 \\
    LASH           &   2    &       2        &   2    &         2       &   3    &       3          &   3    & 3 \\
    UPDN           &   1    &       1        &   1    &         1       &   1    &       1          &   1    & 1 \\
    DFSSSP         &   8    &       8        &   8    &         8       &   8    &       8          &   8    & 8 \\
    SSSP-DF        &   3    &       3        &   4    &         3       &   5    &       3          &   6    & 3 \\
    MINHOP-DF      &   4    &       3        &   5    &         3       &   6    &       3          &   6    & 3 \\ \hline
    \end{tabular}%
}
\label{t:scenarios}
\end{table}

\begin{figure}[!hptb]
        \centering
        \includegraphics[width=\textwidth]{sim-keys}
        \vspace{-.8 cm}

         \subfloat[72-nodes \dfly{} (a4h2p2)]{\includegraphics[width=0.45\textwidth]{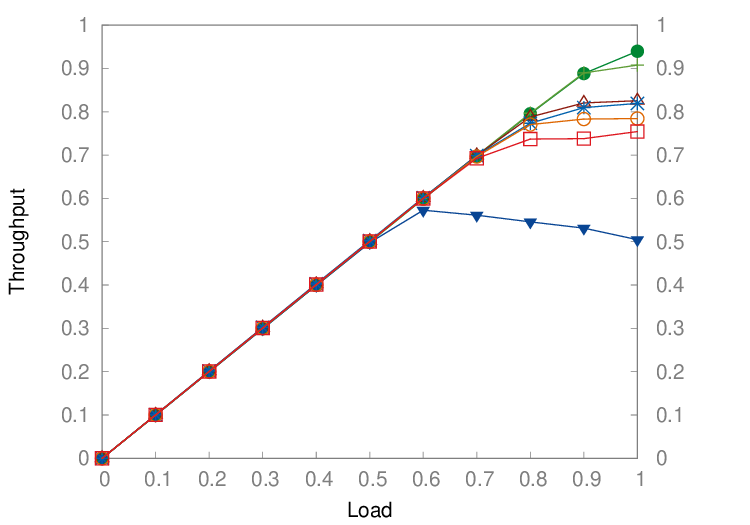}
                 \label{f:r72n}}
        \subfloat[342-nodes \dfly{} (a6h3p3)]{\includegraphics[width=0.45\textwidth]{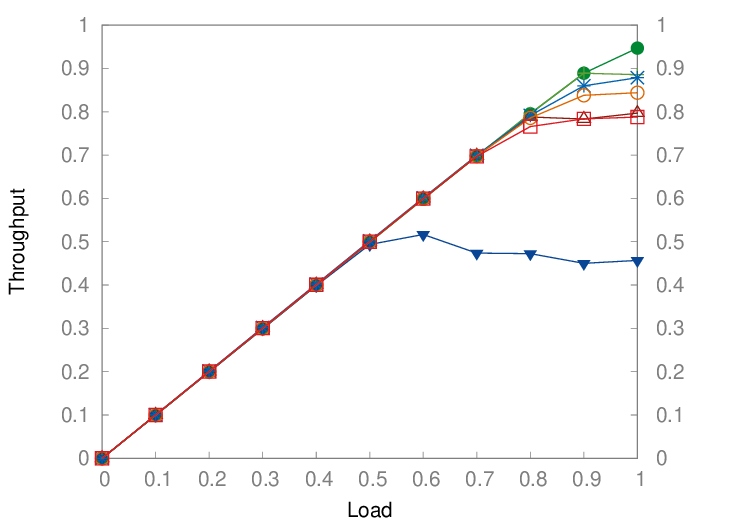}
                \label{f:r342n}}

        \subfloat[1056-nodes \dfly{} (a8h4p4)]{\includegraphics[width=0.45\textwidth]{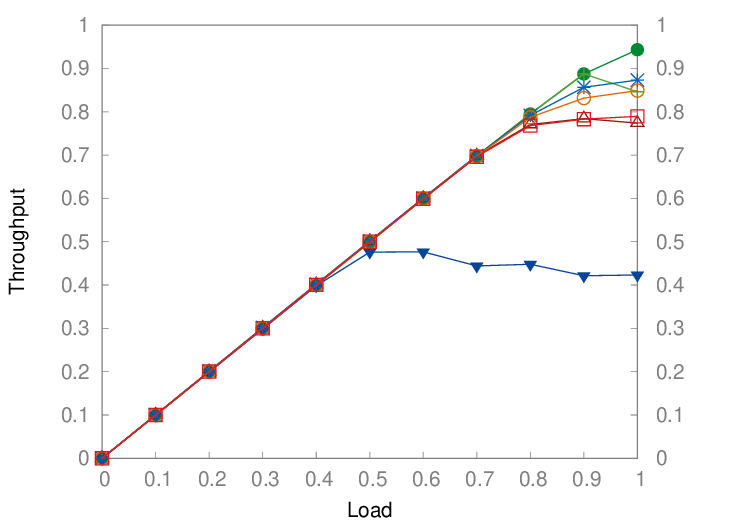}
                \label{f:r1056n}}
        \subfloat[2550-nodes \dfly{} (a10h5p5)]{\includegraphics[width=0.45\textwidth]{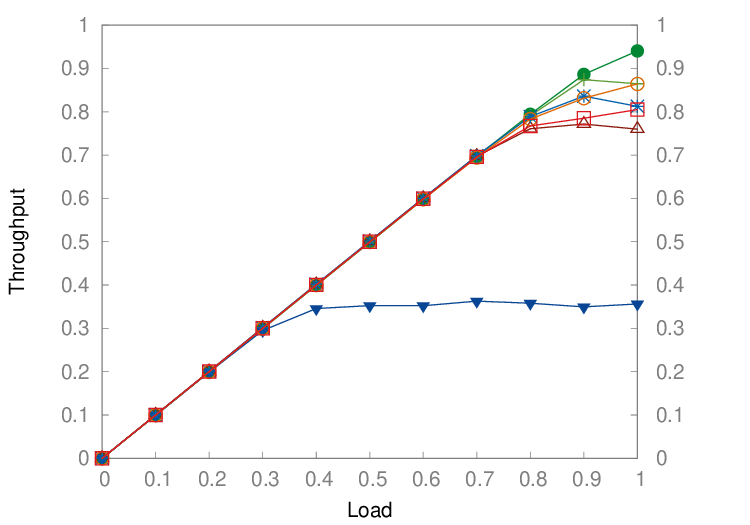}
                \label{f:r2550n}}

        \caption{Balanced fully-connected \dflys{} under random uniform traffic pattern (VOQ is used).}

        \label{f:sim-bal-uniform}
\end{figure}

\begin{figure}[!hptb]
        \centering
        \includegraphics[width=\textwidth]{sim-keys}
        \vspace{-.8 cm}

         \subfloat[72-nodes \dfly{} (a4h2p2)]{\includegraphics[width=0.45\textwidth]{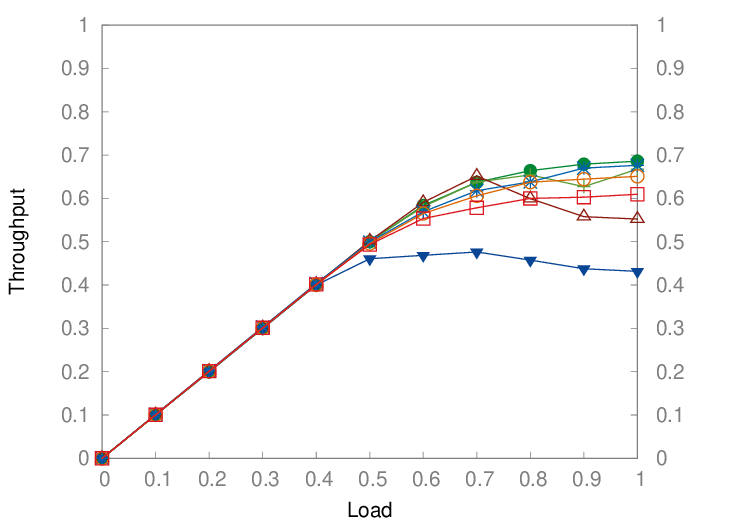}
                 \label{f:s72n}}
        \subfloat[342-nodes \dfly{} (a6h3p3)]{\includegraphics[width=0.45\textwidth]{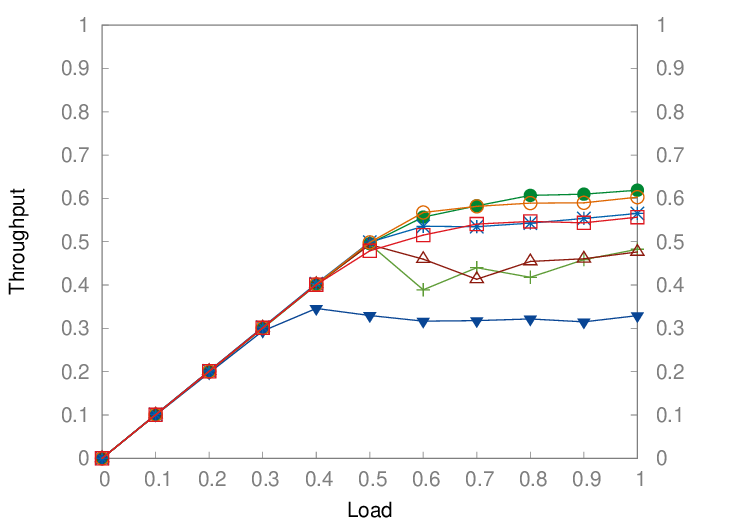}
                \label{f:s342n}}

        \subfloat[1056-nodes \dfly{} (a8h4p4)]{\includegraphics[width=0.45\textwidth]{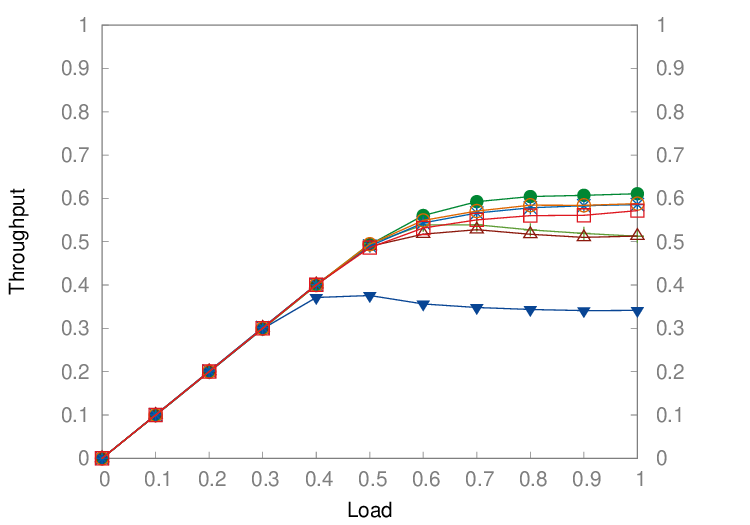}
                \label{f:s1056n}}
        \subfloat[2550-nodes \dfly{} (a10h5p5)]{\includegraphics[width=0.45\textwidth]{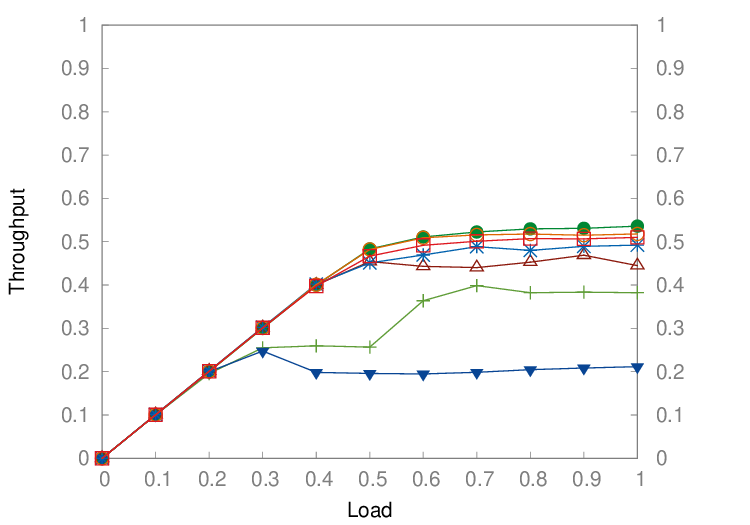}
                \label{f:s2550n}}

        \caption{Balanced fully-connected \dflys{} under 6-point 3D stencil traffic pattern (VOQ is used).}

        \label{f:sim-bal-st3d}
\end{figure}

\begin{figure}[!hptb]
        \centering
        \includegraphics[width=\textwidth]{sim-keys}
        \vspace{-.8 cm}

         \subfloat[72-nodes \dfly{} (a4h2p2)]{\includegraphics[width=0.45\textwidth]{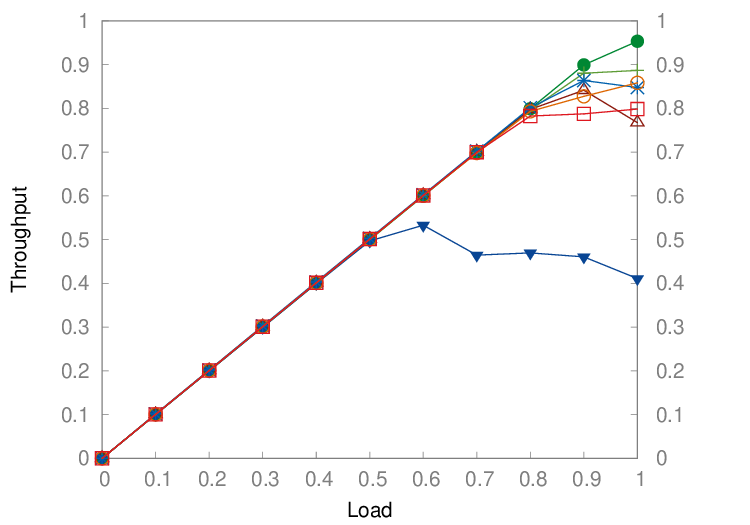}
                 \label{f:h72n}}
        \subfloat[342-nodes \dfly{} (a6h3p3)]{\includegraphics[width=0.45\textwidth]{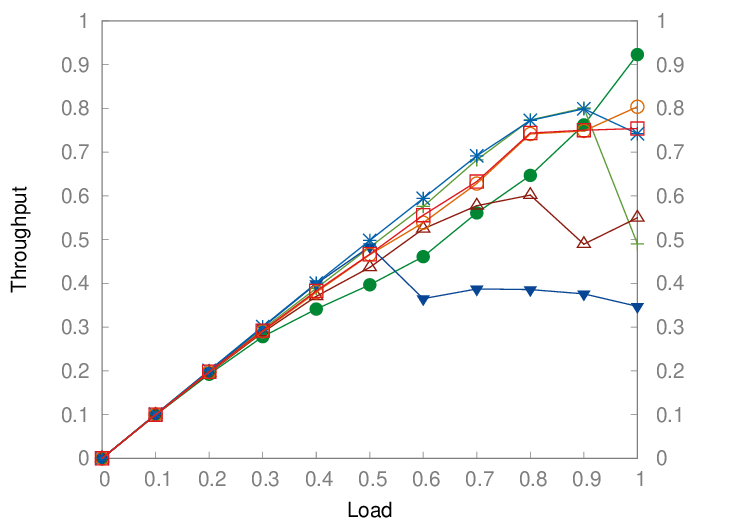}
                \label{f:h342n}}

        \subfloat[1056-nodes \dfly{} (a8h4p4)]{\includegraphics[width=0.45\textwidth]{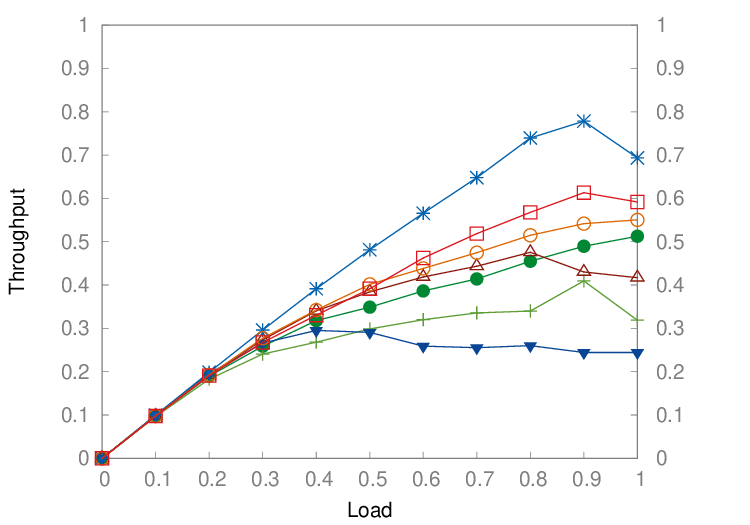}
                \label{f:h1056n}}
        \subfloat[2550-nodes \dfly{} (a10h5p5)]{\includegraphics[width=0.45\textwidth]{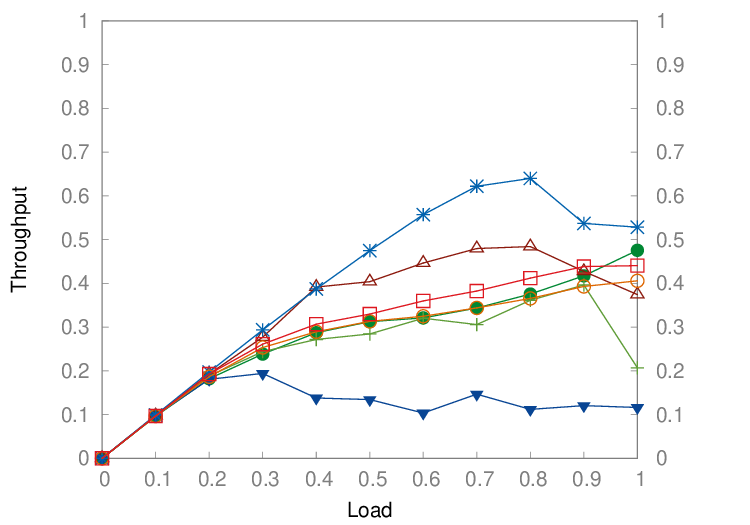}
                \label{f:h2550n}}

        \caption{Balanced fully-connected \dflys{} under hot-spot traffic pattern (VOQ is used).}

        \label{f:sim-bal-hspt1}
\end{figure}

\begin{figure}[!hptb]
        \centering
        \includegraphics[width=\textwidth]{sim-keys}
        \vspace{-.8 cm}

         \subfloat[144-nodes \dfly{} (a4h2p4)]{\includegraphics[width=0.45\textwidth]{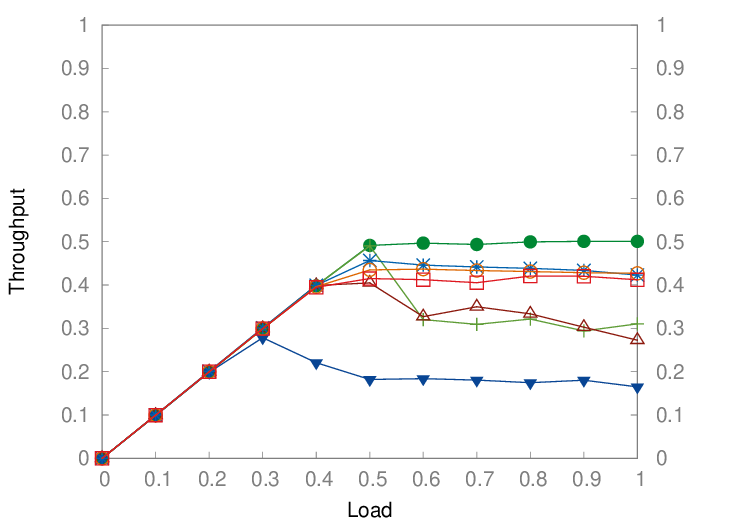}
                 \label{f:r144n}}
        \subfloat[684-nodes \dfly{} (a6h3p6)]{\includegraphics[width=0.45\textwidth]{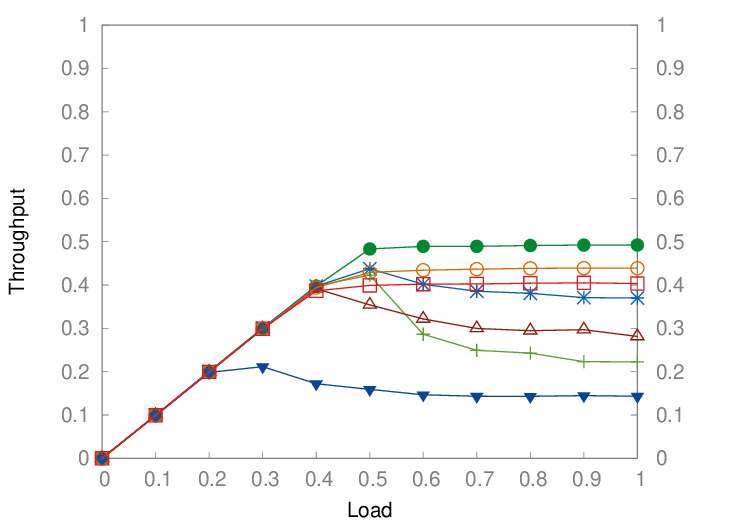}
                \label{f:r684n}}

        \subfloat[2112-nodes \dfly{} (a8h4p8)]{\includegraphics[width=0.45\textwidth]{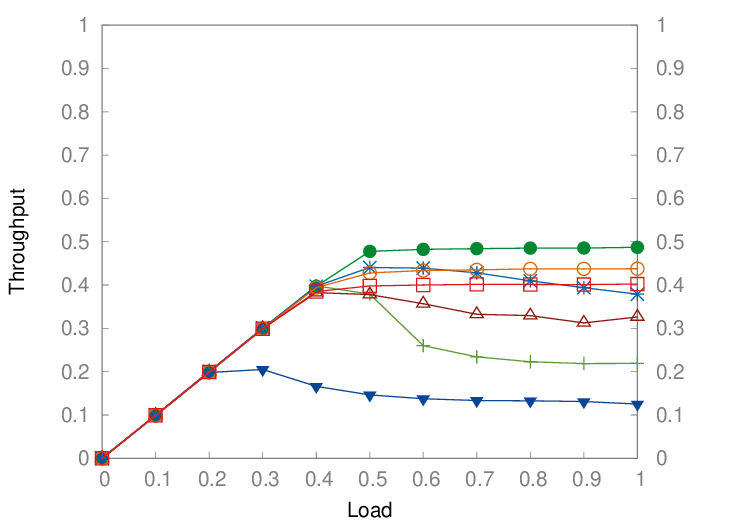}
                \label{f:r2112n}}
        \subfloat[5100-nodes \dfly{} (a10h5p10)]{\includegraphics[width=0.45\textwidth]{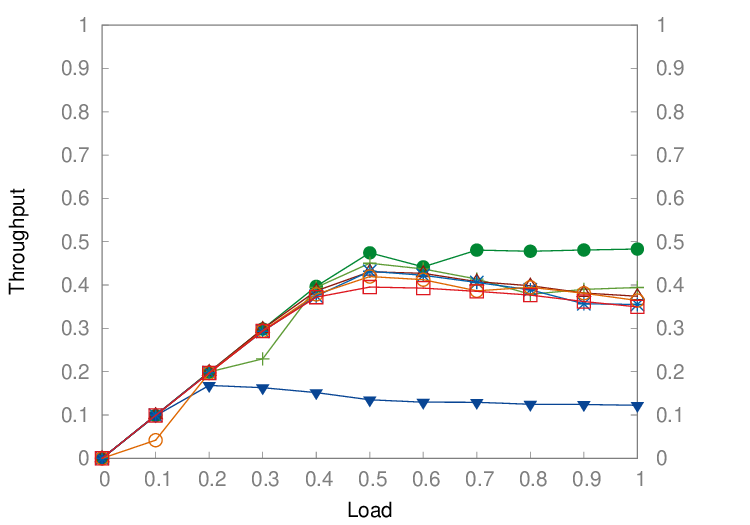}
                \label{f:r5100n}}

        \caption{Oversubscribed fully-connected \dflys{} under random uniform traffic pattern (VOQ is used).}

        \label{f:sim-unbal-uniform}
\end{figure}

\begin{figure}[!hptb]
        \centering
        \includegraphics[width=\textwidth]{sim-keys}
        \vspace{-.8 cm}

         \subfloat[144-nodes \dfly{} (a4h2p4)]{\includegraphics[width=0.45\textwidth]{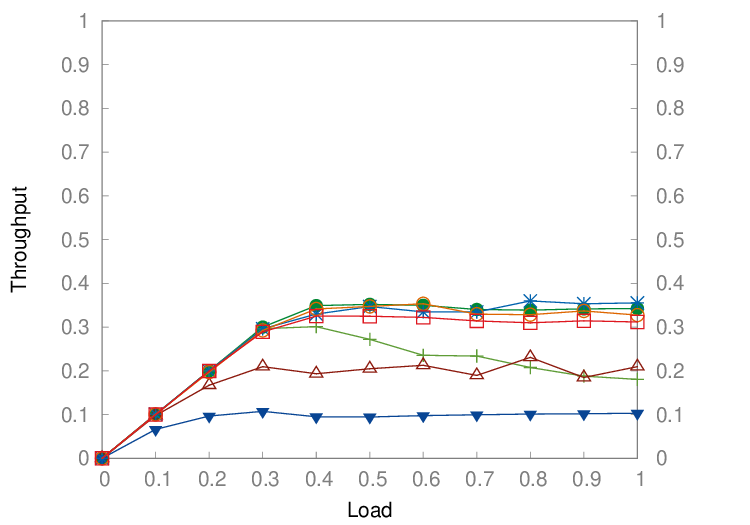}
                 \label{f:s144n}}
        \subfloat[684-nodes \dfly{} (a6h3p6)]{\includegraphics[width=0.45\textwidth]{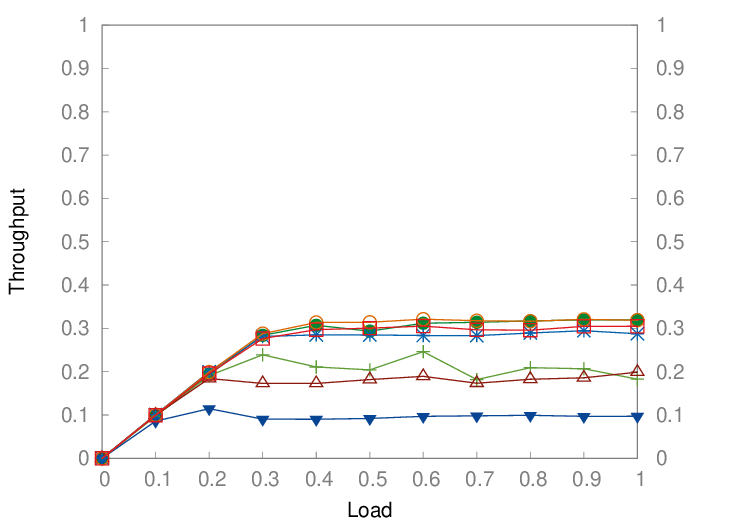}
                \label{f:s684n}}

        \subfloat[2112-nodes \dfly{} (a8h4p8)]{\includegraphics[width=0.45\textwidth]{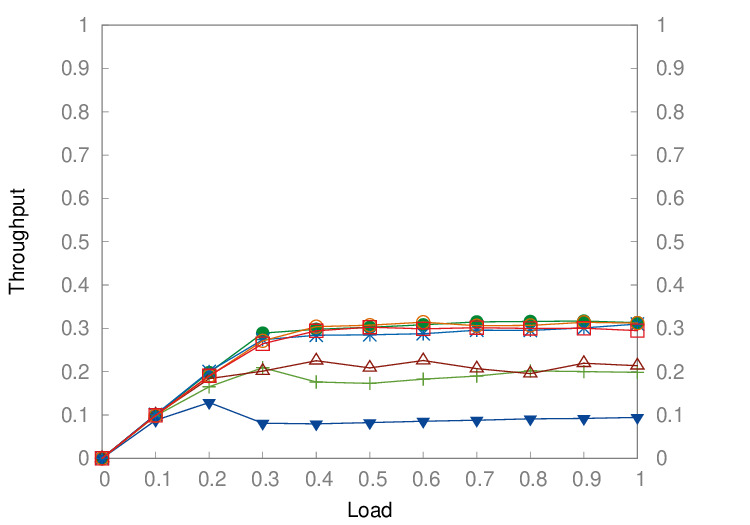}
                \label{f:s2112n}}
        \subfloat[5100-nodes \dfly{} (a10h5p10)]{\includegraphics[width=0.45\textwidth]{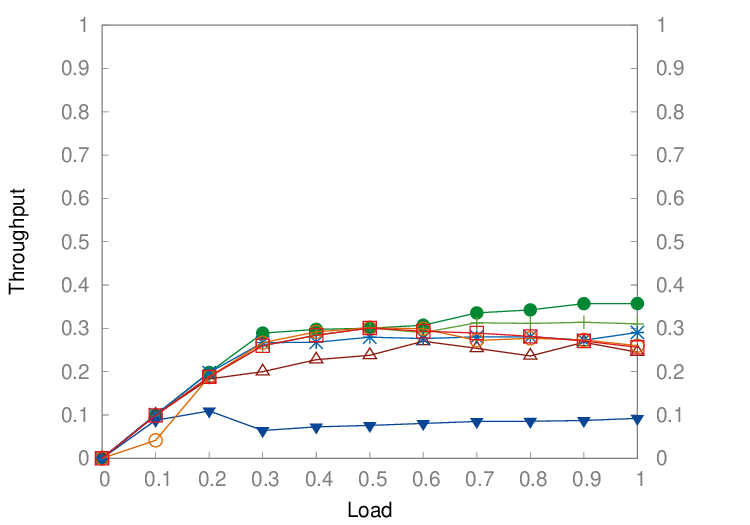}
                \label{f:s5100n}}

        \caption{Oversubscribed fully-connected \dflys{} under 6-point 3D stencil traffic pattern (VOQ is used).}

        \label{f:sim-unbal-st3d}
\end{figure}

\begin{figure}[!hptb]
        \centering
        \includegraphics[width=\textwidth]{sim-keys}
        \vspace{-.8 cm}

         \subfloat[144-nodes \dfly{} (a4h2p4)]{\includegraphics[width=0.45\textwidth]{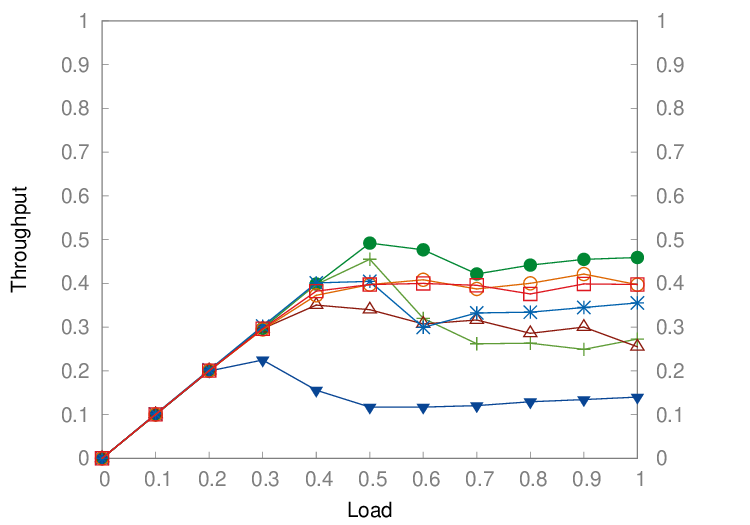}
                 \label{f:h144n}}
        \subfloat[684-nodes \dfly{} (a6h3p6)]{\includegraphics[width=0.45\textwidth]{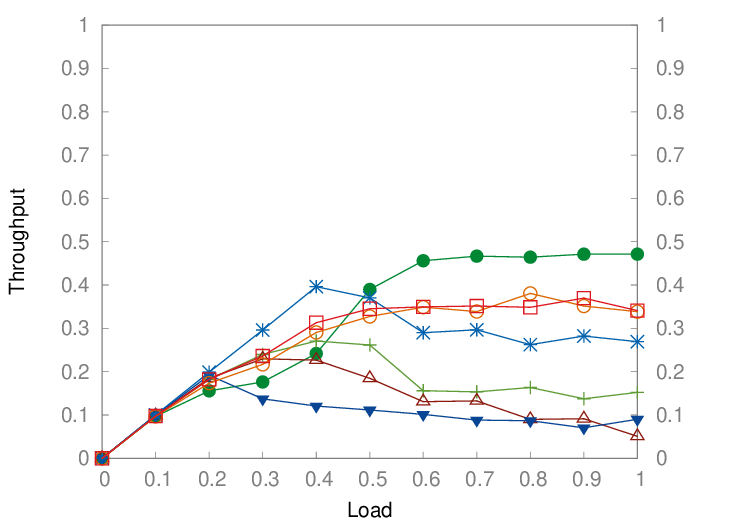}
                \label{f:h684n}}

        \subfloat[2112-nodes \dfly{} (a8h4p8)]{\includegraphics[width=0.45\textwidth]{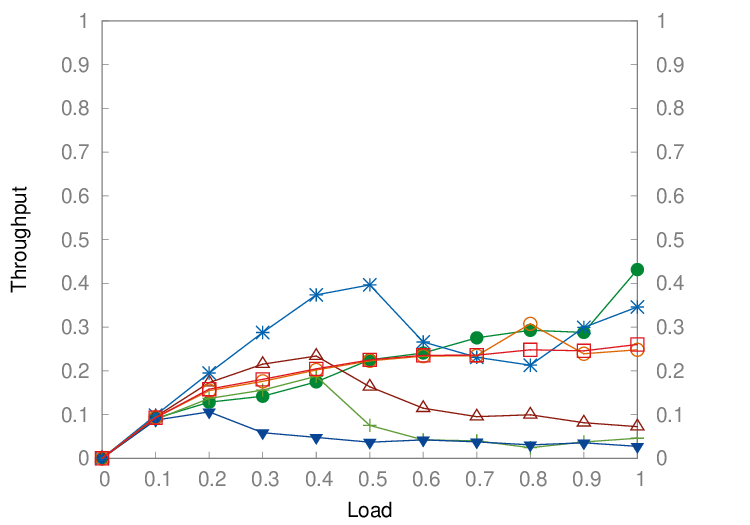}
                \label{f:h2112n}}
        \subfloat[5100-nodes \dfly{} (a10h5p10)]{\includegraphics[width=0.45\textwidth]{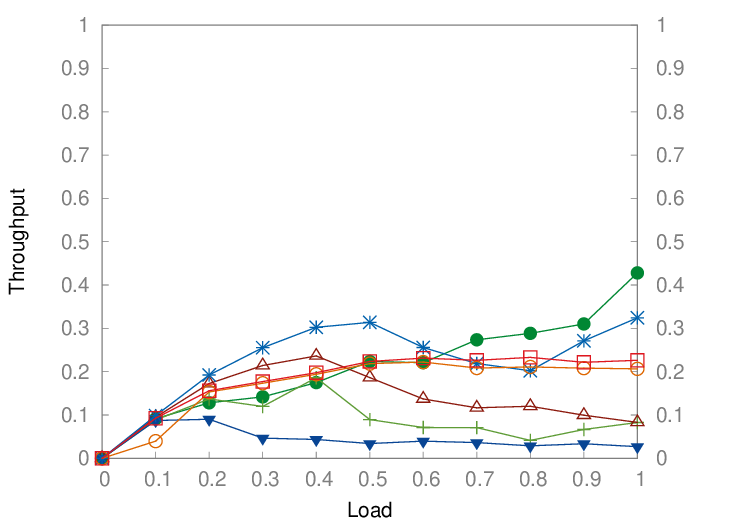}
                \label{f:h5100n}}

        \caption{Oversubscribed fully-connected \dflys{} under hot-spot traffic pattern (VOQ is used).}

        \label{f:sim-unbal-hspt1}
\end{figure}

Figures \ref{f:sim-bal-uniform}, \ref{f:sim-bal-st3d} and \ref{f:sim-bal-hspt1} depict the performance results obtained
for the routing engines when applied to the \dfly{} networks of Table \ref{t:scenarios} under respectively uniform, 6-point 3D stencil, and host-spot traffic patterns.
Specifically, these figures show the accepted traffic (throughput) normalized against the maximum link bandwidth, as a function of traffic load.
Figure \ref{f:sim-unbal-uniform}, \ref{f:sim-unbal-st3d} and \ref{f:sim-unbal-hspt1} show simulation results for oversubscribed
(i.e. not balanced, $a = 2h = p$) \dflys{}, where the number of endnodes doubles with respect to the networks of Table \ref{t:scenarios}.


In general, all the routing engines show a stable performance for the different network sizes under random uniform and 3D stencil traffic patterns,
which is coherent with the results shown in Section \ref{s_sim_sw_features}, independently whether the \dflys{} are balanced or not.
The only exception we see is for D3R and LASH, which show unstable performance when used in oversubscribed \dflys{}, because of the intra-group contention generated by the uniform traffic pattern.
By contrast, under hot-spot traffic, when the network size increases all the routing engines exhibit some degree of variability in their performance results, since  \dflys{}
are sensitive to this type of traffic pattern \cite{bhatele_2011}.
However, DLA is able to provide a throughput comparable to routing engines that require a larger number of VLs.

Overall, DLA exhibits the best performance results on most scenarios, since the simulator is configured with enough buffer space and VOQ (see Section \ref{s_sim_sw_features}).
On the other hand, UPDN gets always the worst performance because it configures longer paths,
and favors the appearance of bottlenecks at the root of the directed graphs it is based on.

\subsection{Experiments in a real cluster}

This section shows experiment results for DLA, D3R, LASH, UPDN, DFSSSP, and DF-DN (SSSP-DF and MINHOP-DF) performed under real traffic
workloads in the Cluster for the Evaluation of Low-Latency Architectures (CELLIA) built from \ibl-based
hardware\footnote{CELLIA belongs to the RAAP research group \cite{web-raap-group}, from the Albacete Research Institute of Informatics at the
University of Castilla-La Mancha, Spain.}.
CELLIA allows us to test the correctness of the implementation of a routing engine by comparing simulation results against real-workload execution.
Each server node in CELLIA is a HP Proliant DL120 Gen9 with Intel Xeon E5-2630v3 8-cores 1.80GHz processor and 32GB of RAM memory.
We installed Ubuntu 16.04.3 LTS (Xenial) with a kernel version \texttt{4.4.0 x86\_64}.
Each node has a dual-port Mellanox\texttrademark~ConnectX3 MCX353A-QCBT HCA working at QDR speed (i.e. 40 Gbps throttled to actual 32 Gbps due to the 8b/10b encoding protocol).
HCAs are attached through a x16 PCIe v3.0 interface.
The HCA drivers and firmware are supplied by Mellanox (HCA firmware \texttt{v2.42.5000}).

Specifically, for this evaluation we built a $42$-node ($a=3$, $h=3$, $p=2$) fully-connected \dfly{} topology in CELLIA.
We used $21$ $8$-port Mellanox\texttrademark~IS5022 switches to build a \dfly{} with $7$ groups and $3$ switches per group.
Switches ports also work at QDR speed.
Cables are QSFP Mellanox\texttrademark, suitable for QDR speed.
Both HCAs and Switches offer $9$ Virtual Lanes (VLs) per port: $8$ data VLs  and $1$  management VL.
We run a modified version of OpenSM \texttt{v3.3.19} \cite{opensm} that includes all the routing engines explained and analyzed in the previous sections.

\begin{figure}[!hptb]
        \centering
        \includegraphics[width=\textwidth]{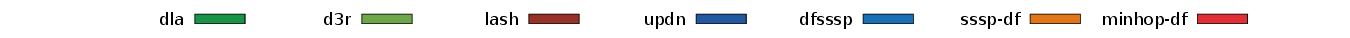}
        \vspace{-.6cm}

        \subfloat[Netgauge (Nto1, 1toN and EBB)]{\includegraphics[width=0.33\textwidth]{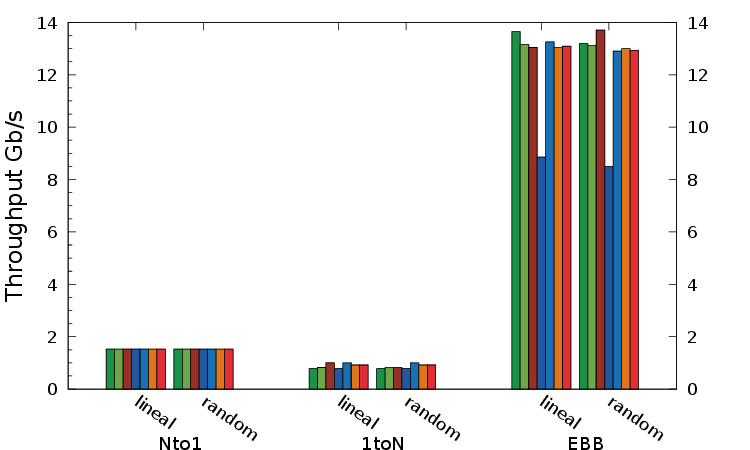} 
                \label{f:ng}}
        \subfloat[Graph500 (Simple and Replicated)]{\includegraphics[width=0.33\textwidth]{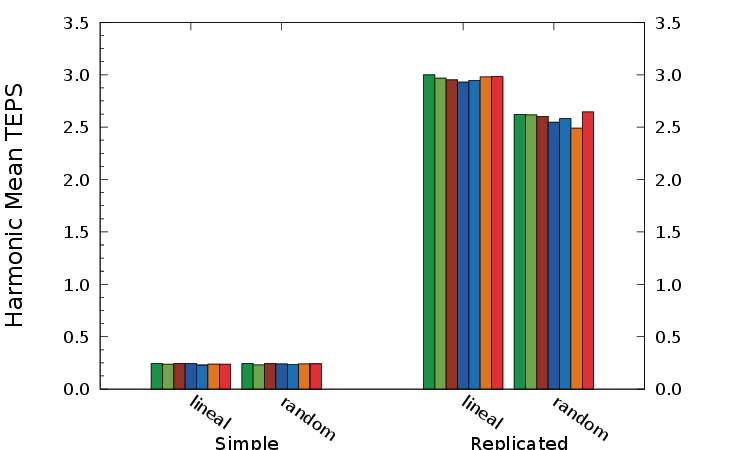}
                \label{f:g500}}
        \subfloat[HPCG (GFLOPs and Execution time)]{\includegraphics[width=0.33\textwidth]{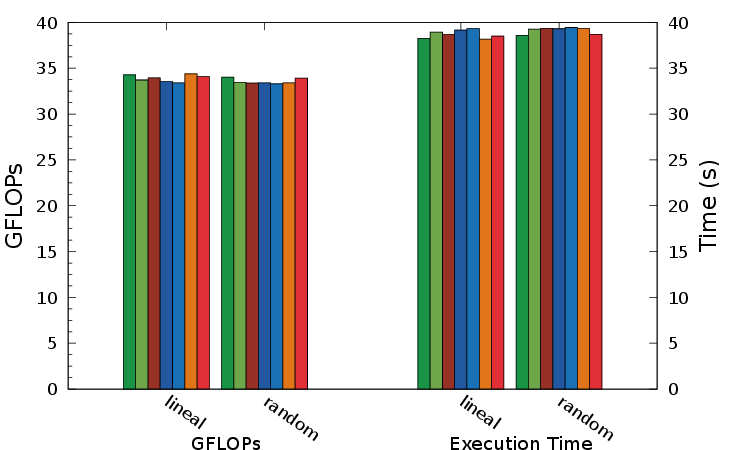}
                \label{f:hpcg}}

        \subfloat[HPCC (Ping Pong)]{\includegraphics[width=0.33\textwidth]{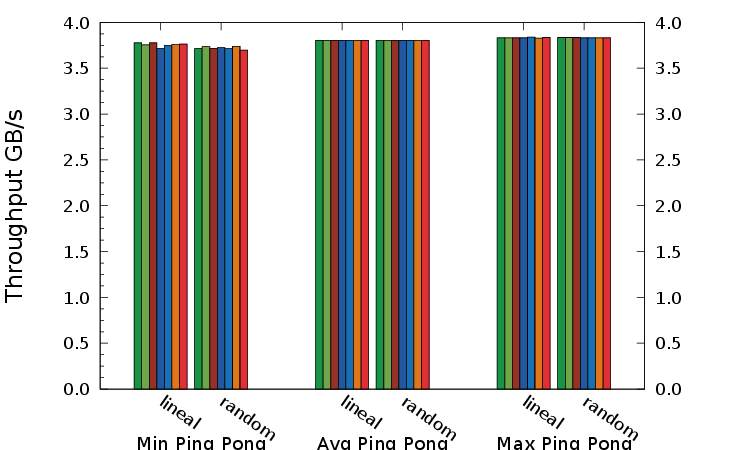}
                \label{f:hpccbw1}}
        \subfloat[HPCC (Ordered Ring) and PTRANS]{\includegraphics[width=0.33\textwidth]{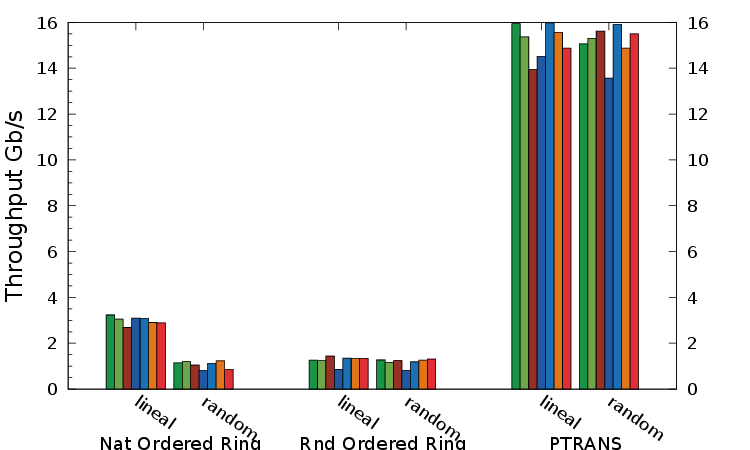}
                \label{f:hpccbw2}}
        \subfloat[HPCC (MPI Random Access)]{\includegraphics[width=0.33\textwidth]{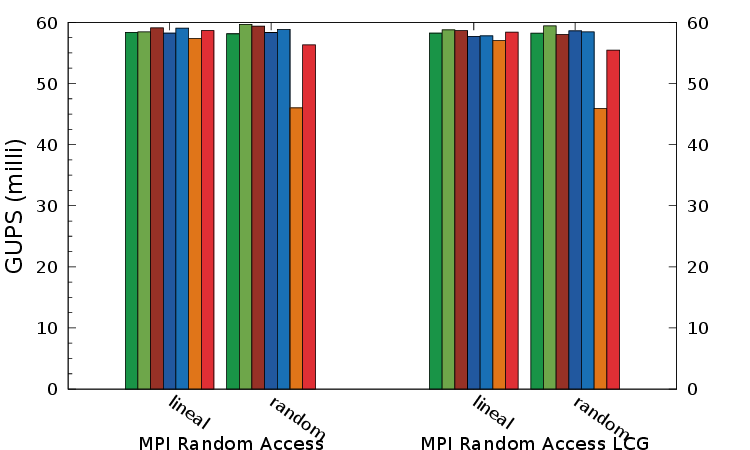}
                \label{f:hpccgups}}

        \subfloat[HPCC (Ping Pong)]{\includegraphics[width=0.33\textwidth]{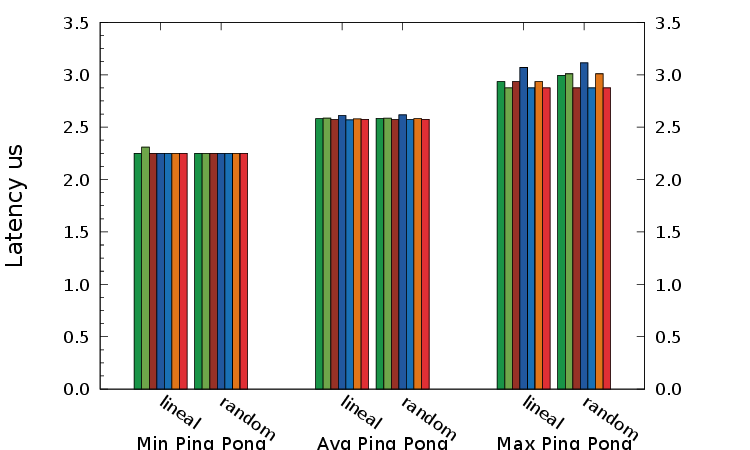}
                \label{f:hpcclat1}}
        \subfloat[HPCC (Ordered Ring)]{\includegraphics[width=0.33\textwidth]{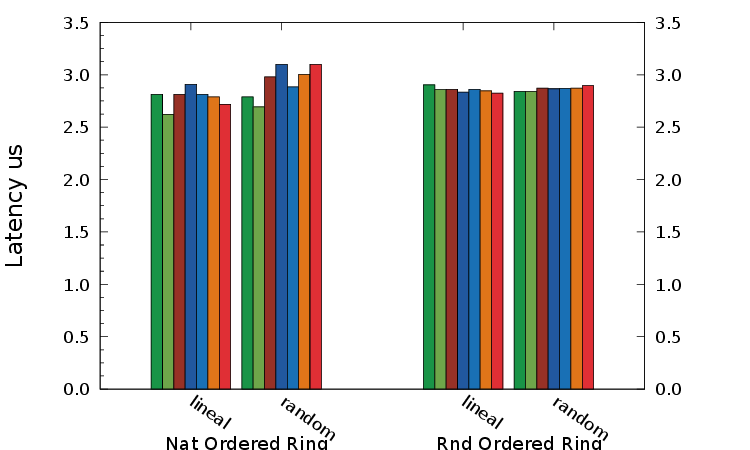}
                \label{f:hpcclat2}}
        \subfloat[NAMD (apoa and f1atpase)]{\includegraphics[width=0.33\textwidth]{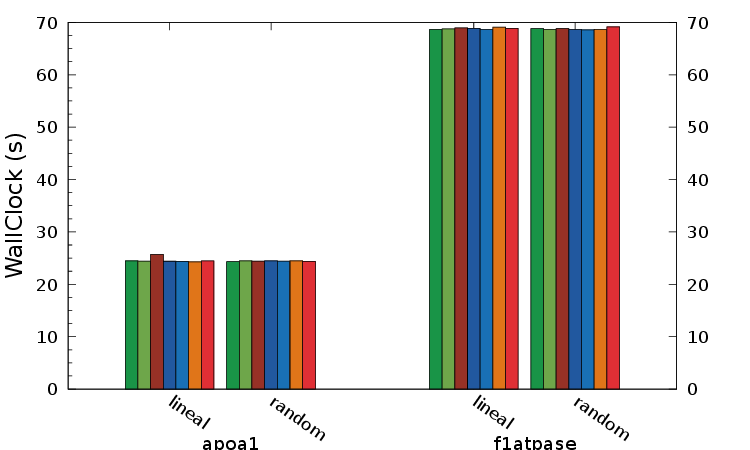}
                \label{f:namd}}

        \caption{Performance results of MPI-based real workloads in the $42$-node \dfly{} configured in the real \ib-based cluster CELLIA.}
        \label{f:benchmarks}
\end{figure}

In order to validate our implementation of DLA, and the simulation results shown in Sections \ref{s_sim_sw_features} and \ref{s_sim-experiments}, we have run a single experiment in CELLIA
for each combination of routing engines (DLA, D3R, LASH, UPDN, DFSSSP, SSSP-DF and MINHOP-DF),  benchmarks
(Netgauge \cite{hoefler-netgauge-hpcc07}, HPCG \cite{hpcg}, HPCC \cite{hpcc}, Graph500 \cite{graph500} and NAMD \cite{namd})
and task mappings (Random and Linear, i.e. the $MPI~task_i$ is assigned to endnode $i$).
Basically, it is the same methodology used to validate our previous routing engine proposal \cite{maglione2018_d3r}.
Figure \ref{f:benchmarks} shows the performance results of these experiments.

Note that the real applications generate one out of four traffic patterns or a combination of them. 
Specifically, HPCC tests Ping-Pong and Ordered-Ring generate an adversarial traffic pattern. Other tests,
such as PTRANS or Netgauge-ebb generate a many-to-many traffic pattern,
while Netgauge Nto1 and Netgauge 1toN generate a hotspot (i.e. many-to-one) and one-to-many pattern, respectively.
The remaining tests generate a combination of the above (although not necessarily a combination of all the patterns)
and we use them to extend the consistency of the evaluation, looking at other real workloads.

In general, there are small variations among all the evaluated routing engines, because of the CELLIA's small size network
(i.e. a 42-node \dfly{} with 21 switches).
Note, however, that the simulation in a small network (see Figure \ref{f:r72n}) and the experiments in CELLIA show qualitatively similar performance results.
For instance, under many-to-many traffic patterns, UPDN shows a performance degradation because of the longer routes it configures and to the root switch becoming a bottleneck.
For DLA, the performance results are similar, in many cases, to those of D3R, DFSSSP and LASH.
The same applies to SSSP-DF and MINHOP-DF, except for the HPCC test \emph{MPI Random Access},
where the performance of both are below the ones of other routing engines.

In summary, experiments under real and simulated scenarios confirm the implementation of DLA as a 
viable routing engine for \ib-based \dflys{}, offering a similar or better performance than other routing engines under different applications
while requiring a very reduced number of resources ($1$ SL and $2$ VLs).

\section{Conclusions}
\label{s_conclusions}
The overall objective of the work described in this paper is to provide the networks based on the InfiniBand architecture (IB) with a suitable implementation of the deadlock-free minimal routing algorithm (DLA)
proposed by Kim and Dally for \dflys{}. Prior to this work, this algorithm had not been implemented as a routing engine in the \ib{} control software (OpenSM), mainly due to the restrictions in the \ib{} specification regarding the shifting of Virtual Lanes (VLs), which DLA requires to guarantee deadlock freedom. In order to fill this gap, we have proposed a straightforward method to overcome these restrictions, so that finally DLA can be used as the routing engine in real \ib-based clusters configured with \dfly{} topology, instead of the generic (i.e. topology agnostic) routing engines previously available in OpenSM.

In more detail, our implementation of DLA is restricted to \dflys{} with fully-connected inter- and intra-group networks (see Section \ref{s_problem_statement}). In addition, an ``asymmetric'' computation of the Service-Level-to-Virtual-Lane (SL2VL) tables is required, which may introduce some degree of complexity in the network configuration. The results obtained both from simulations and from experiments performed in a real \ib-based cluster confirm
that our implementation of DLA is able to outperform other routing engines available in OpenSM in networks based on switches implementing Virtual Output Queuing (VOQ).

Moreover, we have analyzed the requirements of the different routing engines in terms of the number of required SLs and VLs, and we can conclude that DLA requires fewer resources than most of the remaining and already implemented routing engines
included in OpenSM.

\section*{Acknowledgment}
This work has been jointly supported by the Spanish Ministry of Science, Innovation \& Universities under the project RTI2018-098156-B-C52,
Spanish MINECO under project UNCM13-1E-2456, and by Junta de Comunidades de Castilla-La Mancha under the projects POII10-0289-3724, PEII-2014-028-P and SBPLY/17/180501/000498.
German Maglione-Mathey is funded by the Universidad de Castilla-La Mancha (UCLM)  with a pre-doctoral contract PREDUCLM16/29.
Jesus Escudero-Sahuquillo is funded by the Universidad de Castilla-La Mancha (UCLM) and the European Commission (FSE funds),
with a contract for accessing the Spanish System of Science, Technology and Innovation, for the implementation of the UCLM
research program (UCLM resolution date: 31/07/2014).

\section*{References}
\bibliographystyle{elsarticle-num}
\bibliography{BiblioDB}

\end{document}